\documentclass[a4paper,11pt]{article}
\pdfoutput=1
\usepackage{jheppub}

\usepackage{graphicx}
\usepackage{booktabs}
\usepackage{amsmath,amsfonts,amssymb,amsthm}

\usepackage{hyperref}
\usepackage{verbatim}
\usepackage[normalem]{ulem}
\usepackage{xcolor}

\newcommand{\fft}[2]{\frac{#1}{#2}}

\newcommand{\nn}{\nonumber}
\newcommand{\fsl}[1]{{\ooalign{\(#1\)\cr\hidewidth\(/\)\hidewidth\cr}}}
\newcommand*{\pd}[3][]{\ensuremath{\frac{\partial^{#1} #2}{\partial {#3}^{#1}}}}

\DeclareMathOperator{\Tr}{Tr}

\preprint{LITP-25-11}

\title{Quantized Giant Gravitons as the Periodic Table of Supersymmetric States: D3, M2 and M5}

\author{Evan Deddo,}
\author{Sabarenath Jayaprakash,}
\author{James T. Liu}
\author{and Leopoldo A. Pando Zayas}

\emailAdd{evdedd@umich.edu}
\emailAdd{sabare@umich.edu}
\emailAdd{jimliu@umich.edu}
\emailAdd{lpandoz@umich.edu}

\affiliation[]{Leinweber Institute for Theoretical Physics, University of Michigan, Ann Arbor, MI 48109, USA}
\abstract{
    We consider giant gravitons in the probe approximation when they are described by classical brane configurations in AdS$_{p+2}\times S^{q+2}$ wrapping a particular $q$-cycle and spinning in $S^{q+2}$. We quantize the full set of fluctuations of these configurations and show that they are sufficient to capture all the supersymmetric single-letter indices of the corresponding dual field theories. We explicitly discuss the cases of D3, M2 and M5 branes and reproduce the single-letter indices for all fractions of supersymmetry. We also provide a new derivation of the full finite-$N$ half-BPS index by promoting certain fluctuations to matrix-valued fields. We elaborate on the obstructions for the general finite-$N$ computations.  Given that the single-letter partition functions are the building blocks of all supersymmetric enumerations, including for the black hole entropy, our work provides a direct gravitational counting of those degrees of freedom modulo finite-$N$ obstacles due to non-Abelian effects.   
}

\begin{document}

\maketitle

\section{Introduction}

One of the main achievements of the AdS/CFT correspondence has been to provide a microscopic enumeration of the states responsible for the Bekenstein-Hawking entropy of large classes of asymptotically AdS black holes \cite{Benini:2015eyy,Cabo-Bizet:2018ehj, Choi:2018hmj, Benini:2018ywd}. This progress crucially relied on translating the problem of enumeration of microstates to the CFT side. Once outsourced to a field theory question, one needs only compute the appropriate (protected) partition function to obtain the black hole entropy proportional to a quarter of the horizon area \cite{Choi:2019miv, Kantor:2019lfo, Nahmgoong:2019hko, Choi:2019zpz, Nian:2019pxj, Bobev:2019zmz, Benini:2019dyp,Crichigno:2020ouj}. The superconformal index  has been particularly powerful, providing a window into sub-leading and non-perturbative contributions of the degeneracy of states  \cite{GonzalezLezcano:2020yeb,Aharony:2021zkr,Chen:2023lzq,Cabo-Bizet:2023ejm}. This resounding success, however,  does not address the important question of directly enumerating the states in the gravitational picture of the correspondence. A direct gravitational counting,  without reference to the field theory picture, is likely to provide important insights that can be extended beyond strictly supersymmetric zero temperature black holes. {\it In this manuscript we focus on reproducing the results of various enumeration problems by directly quantizing and counting gravitational states arising as excitations of giant gravitons in the probe approximation. }

An important insight into how to potentially identify the gravitational degrees of freedom was indirectly provided by the dual field theory itself. Namely, it was established that in various field theories characterized by a gauge group of rank $N$, superconformal indices  at finite $N$ admit expansions of the schematic form
\cite{Bourdier:2015wda,Arai:2020qaj,Imamura:2021ytr,Gaiotto:2021xce}:
\begin{equation}
  {\cal I}_N(q)={\cal I}_\infty (q) \sum_m q^{mN}\hat{I}_m^{\rm GG}(q),  
\end{equation}
where $q$ is a representative fugacity. The above type of formulas express the finite-$N$ index, ${\cal I}_N(q)$, in terms of corrections to the infinite-$N$ index, ${\cal I}_\infty(q)$, and an expansion involving powers of $q^N$ and  the  giant graviton index $\hat{I}_m^{\rm GG}(q)$. This formula admits a holographic interpretation where ${\cal I}_\infty(q)$ is the supergravity or Kaluza-Klein index and the role of the giant graviton expansion is to correct  the infinite-$N$ counting to a finite-$N$ one, often thought of as the implementation of trace relations.  Holographically, the role of the expansion parameter $q^N$ is anticipated to be played by giant gravitons since they happen to have the appropriate tension proportional to $N$. Our precise goal is to gain insight into $\hat{I}^{
\rm GG}_m(q)$ directly from gravitational degrees of freedom. 

Some progress on the direct counting using gravitational degrees of freedom has been recently  achieved in the context of $\frac{1}{2}$-BPS configurations using the probe description of giant gravitons \cite{Lee:2023iil,Eleftheriou:2023jxr,Lee:2024hef,Eleftheriou:2025lac}. There is a fully back-reacted geometry describing giant gravitons and the authors of \cite{Deddo:2024liu} showed that a semiclassical quantization of such geometries  reproduces the $\frac{1}{2}$-BPS index at finite $N$, including its giant graviton expansion term by term.

In this manuscript we first study the quadratic fluctuations around probe description of giant gravitons. Previous partial results for the fluctuations of giant gravitons as probes in supergravity have been reported as far back as \cite{Das:2000st} and as recently as \cite{Beccaria:2024vfx,Gautason:2024nru}. In this manuscript we perform a complete and systematic analysis of the fluctuations of giant gravitons described by D3 branes in AdS$_5\times S^5$, M2 branes in AdS$_7\times S^4$ and M5 in AdS$_4\times S^7$. After canonically quantizing these fluctuations, we present details of the enumeration of single-letters for all supersymmetric indices, including those that capture black hole microstates.  

Beyond the single-brane analysis, one would like to tackle the fluctuations of multiple giant gravitons. At this level the situation is more complicated, requiring both stacks of coincident giants as well as intersecting orthogonal giants. The obstructions we face are ultimately rooted in (1) the lack of a general non-Abelian DBI action and (2) the lack of a description of modes on brane intersections. In general we are limited to a single-brane analysis, which prevents us from obtaining a full finite-$N$ result directly relevant for the black hole entropy. However, there are some results for coincident giant configurations, especially in the case of the $\frac{1}{2}$-BPS index in $\mathcal{N}=4$ SYM. The $\frac{1}{2}$-BPS index involves only a single stack of branes, and simple generalizations of the DBI action are sufficient to describe all required fluctuations. The full giant graviton expansion has previously been achieved via a certain localization argument \cite{Eleftheriou:2023jxr,Lee:2024hef, Eleftheriou:2025lac}. We provide an alternative derivation of the $\frac12$-BPS index by promoting the fluctuation coordinates to matrix-valued fields. This strategy, however, is not applicable in more general forms of the superconformal index. {\it We have the atoms but cannot construct the molecules in the general case.}

It is worth clarifying the origin of the various enumerations within the bigger framework of the AdS/CFT correspondence and string theory at large. We have three different enumerations originating in: (i)  directly in ${\cal N}=4$ SYM, (ii) the fully back-reacted gravitational theory and (iii) as states  living in the world-volume of giant gravitons. The origins of enumerations (i) and (ii) can be traced to the two dual forms of viewing, for example, a D3 brane; either in its open string description  or in its closed string description. Giant gravitons are at the center of description (iii) which can be interpreted as an alternative open string description. Similar ideas have recently been advanced in a related context by \cite{Gopakumar:2024jfq} and our work can be viewed as explicitly fleshing out this interesting open-closed-open triality.

The rest of the manuscript is organized as follows. In Section~\ref{Sec:Scalars} we present a unified treatment of scalar fluctuations in a generic supergravity background with metric AdS$_{p+2}\times S^{q+2}$ supported by a $(q+2)$-form flux. We will  further use these results in various sections as ingredients in the full spectrum of excitations of giant graviton branes. Section~\ref{Sec:D3GG} discusses the prototypical example of a giant graviton in the probe approximation;  we present the full spectrum of excitations including the scalars as well as the vector and fermionic excitations of a giant graviton D3 brane in the AdS$_5\times S^5$ IIB supergravity background. In section~\ref{Sec:PartFn} we use the spectrum of excitations of  the giant graviton as a gravitational brane configuration to reproduce all the single-letter indices of ${\cal N}=4$ supersymmetric Yang-Mills.  In section~\ref{Sec:M5}, we present the spectrum of excitations of the M5 brane, including the anti-symmetric tensor field and the fermionic part of the spectrum; we further present how to recover various indices using a subset of the excitations. Section~\ref{Sec:M2} reproduces the spectrum and some of the indices for the M2 brane. Section~\ref{Sec:nonAb} reproduces the full giant graviton expansion of the $\frac{1}{2}$-BPS index by generalizing the analysis of a single brane to a stack of multiple D3 branes. We conclude in Section~\ref{Sec:Conclusions} and relegate spinor conventions  and some useful character formulas to two appendices.

%%%%%%%%%%%%%%%%%%%%%%%%%%%%%%%%%%%%%%%%%%%%%%%%%%%%%%%%%%%%%%%
\section{Scalar fluctuations of a giant graviton in general dimensions}\label{Sec:Scalars}

In general, a giant graviton can be described as a brane configuration in a given supergravity background with metric $AdS_{p+2}\times S^{q+2}$. The defining property of the giant graviton is its  angular momentum on the sphere part of the background and that it wraps some cycle on the sphere with  world-volume typically $\mathbb{R}\times S^q$. In this section we will discuss the general form of scalar fluctuations; they form part of a fairly  universal sector in every configuration and can be treated in a unified fashion.

Consider a supergravity solution of the form $AdS_{p+2}\times S^{q+2}$ supported by $q+2$ form flux:
\begin{align}
    ds^2&=-(1+r^2/\tilde{L}^2)dt^2+\fft{dr^2}{1+r^2/\tilde{L}^2}+r^2 d\Omega_p^2\nn\\
    &\qquad+L^2\left(dx^2+dy^2+\fft{(xdx+ydy)^2}{1-x^2-y^2}+(1-x^2-y^2)d\Omega_q^2\right),\nn\\
    \bar A_{q+1}&=L^{q+1}\fft{(1-x^2-y^2)^\fft{q+1}{2}-1}{x^2+y^2}(xdy-ydx)\wedge d\Omega_q^2.
\label{eq:AdSS}
\end{align}
Note that we have written the $S^{q+2}$ metric in a convenient form for the giant graviton.  In particular, we take the giant graviton to be a $q$-brane wrapping $S^q \subset S^{q+2}$.  Maximal giants will be located at the `North pole' of the $S^{q+2}$, so we have used a non-singular coordinate patch with $x$ and $y$ parametrizing that region.  The bosonic part of the brane action on this background is
\begin{equation}
    S_B=-T_q\int d^{q+1}\xi\sqrt{-g}+T_q\int A_{q+1},
    \label{eq:DBI}
\end{equation}
where $g_{\alpha\beta}$ and $A_{q+1}$ are pull backs to the world-volume of the brane. At linearized order the scalar and vector fluctuations decouple, so they can be treated separately.

For the scalars, we choose a physical gauge corresponding to a brane wrapped on $S^q\subset S^{q+2}$. Let $\xi^0=t$ and let $\xi^1,\ldots\xi^q$ coorespond to the coordinates on the $S^q\subset S^{q+2}$. The maximal giant corresponds to taking $r=\eta_i=x=y=0$ where the $\eta_i$ with $i=1,\ldots p$ are coordinates on the $S^p\subset \mathrm{AdS}_{p+2}$.  However, we allow for fluctuations by taking
\begin{equation}
    r(\xi^\alpha),\qquad \eta_i(\xi^\alpha),\qquad x(\xi^\alpha),\qquad y(\xi^\alpha).
\end{equation}
Expanding the action (\ref{eq:DBI}) to quadratic order in the scalar fluctuations then gives
\begin{align}
    S&=\fft{N}{L^{q+1}\mathrm{vol}(S^q)}\int d^{q+1}\xi L^q\sqrt{\gamma}\biggl(-1-\fft1{2\tilde{L}^2}r^2+\fft{q}2(x^2+y^2)\nn\\
    &\kern11em+\fft12L^2(\dot x^2+\dot y^2)-\fft12(|\vec\nabla x|^2+|\vec\nabla y|^2)-\fft{q+1}2 L(x\dot y-y\dot x)\nn\\
    &\kern11em+\fft12\dot r^2+\fft12r^2\fft{d\Omega_p^2}{dt^2} -\fft1{2L^2}|\vec\nabla r|^2-\fft{r^2}{2L^2}\;\gamma^{ij}\fft{d\Omega_p^2}{d\xi^i d\xi^j}\biggr).
\end{align}
Here we have used the relation $T_q=N/L^{q+1}\mathrm{vol}(S^q)$ for the D-brane/M-brane tension.
In addition, $\gamma_{ij}$ is the metric on the unit $S^q$, and we have used the notation $|\vec\nabla\varphi|^2=\gamma^{ij}\partial_i\varphi\partial_j\varphi$. This is essentially a standard quadratic scalar Lagrangian. 

However, note that the potential term
\begin{equation}
    -U=1+\fft1{2L^2}r^2-\fft q2(x^2+y^2)\approx(1+r^2/L^2)^{1/2}(1-x^2-y^2)^{q/2},
\end{equation}
corresponds to the volume of the spherical brane (including the redshift factor from the time-component of the metric), and the mixed term containing $(x\dot y-y\dot x)$ arises from the Wess-Zumino coupling in the brane action.

It is convenient to convert the transverse fluctions in AdS$_{p+2}$ from spherical coordinates $r,\eta_1,\ldots,\eta_p$ to rectangular coordinates $\zeta_1,\ldots, \zeta_{p+1}$.
% \begin{equation}
%     \zeta_1+i\zeta_2=r\cos\eta_1e^{i\eta_2},\qquad\zeta_3+i\zeta_4=r\sin\eta_1e^{i\eta_3}.
% \end{equation}
%
We can also uniformize the notation by taking
\begin{equation}
    \zeta_{p+2}=Lx,\qquad\zeta_{p+3}=Ly.
\end{equation}
The scalar action then takes the form
\begin{align}
    S=\fft{N}{L\mathrm{vol}(S^q)}\int d^{q+1}\xi\sqrt\gamma\bigg[-1+\sum_{a=1}^{p+3}\Big(\fft12\dot\zeta_a^2&-\fft1{2L^2}|\vec\nabla \zeta_a|^2-\fft12M_a^2\zeta_a^2\Big)\nn\\
    &-\fft{q+1}{2L}(\zeta_{p+2}\dot\zeta_{p+3}-\zeta_{p+3}\dot\zeta_{p+2})\bigg],
\end{align}
where
\begin{equation}
    M_a^2L^2=
    \begin{cases}
    \begin{array}{rll}
         &L^2/\tilde L^2, &\quad a=1,\ldots p+1\\
        &-q, &\quad a= p+2\text{ or
 } p+3.
    \end{array}
    \end{cases}
\end{equation}
We now expand the scalars $\zeta_a(\xi^\alpha)$ in $S^q$ harmonics
\begin{equation}\label{eq:SH_decomp}
    \zeta_a(\xi^\alpha)=\sum_{\ell,m_j}\zeta_{a,(\ell,m_j)}(t)Y_{\ell,m_j}(\xi^i),
\end{equation}
with
\begin{equation}
    \vec\nabla^2Y_{\ell,m_j}=-\ell(\ell+q-1)Y_{\ell,m_j},
\end{equation}
where $j=1\ldots\lfloor q/2\rfloor$. Note that scalar harmonics $Y_{\ell,m_j}$ transform in the rank $\ell$ symmetric tensor representation of $SO(q+1)$. 
We shall use the convention $Y_{\ell,m_j}^*=Y_{\ell,-m_j}$, with reality of $\zeta_a$ implying that $\zeta_{a,(\ell,m_j)}^* = \zeta_{a,(\ell,-m_j)}$. Integrating over $S^q$ and making use of orthonormality of the spherical harmonics, we find
\begin{align}\label{eq:S}
    S&=\fft{N}L\int dt\biggl[-1+\sum_{\ell,m_j}\biggl(\sum_{a=1}^{p+3}\biggl(\fft12\dot\zeta_{a,(\ell,m_j)}\dot\zeta_{a,(\ell,-m_j)}\nn\\
    &\kern13em-\fft1{2L^2}(M_a^2L^2+\ell(\ell+q-1))\zeta_{a,(\ell,m_j)}\zeta_{a,(\ell,-m_j)}\biggr)\nn\\
    &\kern10em-\fft{q+1}{2L}\left(\zeta_{p+2,(\ell,m_j)}\dot\zeta_{p+3,(\ell,-m_j)}-\zeta_{p+3,(\ell,m_j)}\dot\zeta_{p+2,(\ell,-m_j)}\right)\biggr)\biggr].
\end{align}
Note that the D3 in AdS$_5\times S^5$, M5 in AdS$_4\times S^7$, and M2 in AdS$_7\times S^4$ all satisfy the relation $L/\tilde{L}=(q-1)/2$. Therefore, the mass squared of the AdS fluctuations $\zeta_1,\ldots,\zeta_{p+1}$ is
\begin{equation}
    \left(\fft{L^2}{\tilde{L}^2}+\ell(\ell+q-1)\right)=\left(\ell+\fft{q-1}2\right)^2.
\end{equation}
(For general AdS$\times S$ solutions, $L/\tilde{L}=(q+1)/(p+1)$, \cite{Das:2000st}, and the three cases considered in this paper are in fact the only $(p,q)$ that give integer masses.) Before proceeding with canonical quantization, we find it convenient to scale the fields, $\zeta_a\to\zeta_a/\sqrt{N}$.  In this case, the canonical momenta are
\begin{equation}
    p_{a,(\ell,m_j)}=\begin{cases}\fft1L\dot\zeta_{a,(\ell,-m_j)},&a=1,\ldots,p+1,\nn\\
    \fft1L\dot\zeta_{p+2,(\ell,-m_j)}+\fft{q+1}{2L^2}\zeta_{p+3,(\ell,-m_j)},&a=p+2,\nn\\
   \fft1L\dot\zeta_{p+3,(\ell,-m_j)}-\fft{q+1}{2L^2}\zeta_{p+2,(\ell,-m_j)},&a=p+3,\end{cases}
\end{equation}
and the Hamiltonian is
\begin{align}\label{eq:scalarH}
    H&=\fft1L\biggl[N+\sum_{\ell,m_j}\biggl(\sum_{a=1}^{p+3}\biggl(\fft{L^2}2p_{a,(\ell,m_j)}p_{a,(\ell,-m_j)}+\fft1{2L^2}\left(\ell+\fft{q-1}2\right)^2\zeta_{a,(\ell,m_j)}\zeta_{a,(\ell,-m_j)}\biggr)\nn\\
    &\kern7em+\fft{q+1}2\left(\zeta_{p+2,(\ell,m_j)}p_{p+3,(\ell,m_j)}-\zeta_{p+3,(\ell,m_j)}p_{p+2,(\ell,m_j)}\right)\biggr)\biggr].
\end{align}
Note that all fields have the same mass, and the second line contains the angular momentum generator in the $x$-$y$ plane in our parametrization of $S^{q+2}$:
\begin{equation}\label{eq:scalarR1}
    R_1=\sum_{\ell,m_j}\left(\zeta_{p+2,(\ell,m_j)}p_{p+3,(\ell,m_j)}-\zeta_{p+3,(\ell,m_j)}p_{p+2,(\ell,m_j)}\right).
\end{equation}
Moreover, $R_1$ commutes with the Hamiltonian.  For the AdS$_{p+2}$ fluctuations $(\zeta_1,\ldots,\zeta_{p+1})$, this Hamiltonian is just that of a set of harmonic oscillators.  For $\zeta_{p+2}$ and $\zeta_{p+3}$, we have instead a charged two-dimensional harmonic oscillator in a magnetic field.

The remaining $R$ charges are associated with rotational symmetries along the world-volume of the giant. Under a small rotation by angle $\varepsilon$ in the $i$-th rotation plane, the $S^q$ spherical harmonics transform as
\begin{equation}
    Y_{\ell,m_j}(\xi)\to e^{i m_i \varepsilon}Y_{\ell,m_j}(\xi),
\end{equation}
where the phase factor contains the  $m$ associated with the choice of rotation plane. The extra phase factor may be absorbed into $\zeta_{a(\ell,m_j)}$ in \eqref{eq:SH_decomp} so that
\begin{equation}
    \zeta_{a,(\ell,m_j)}\to e^{i m_i \varepsilon}\zeta_{a,(\ell,m_j)} \approx \left(1+im_i\varepsilon\right)\zeta_{a,(\ell,m_j)}.
\end{equation}
The remaining $R_2\ldots R_q$ are the the Noether charges associated with these symmetries.
\begin{equation}\label{eq:scalarRi}
    R_{i+1}=-i\sum_{\ell,m_j}\sum_{a=1}^{p+3}m_i \zeta_{a,(\ell,m_j)}p_{a,(\ell,m_j)}, \qquad i=1,\ldots,q-1.
\end{equation}
The Hamiltonian can be quantized in the usual manner by introducing

\begin{align}
    \zeta_{a,(\ell,m_j)}&=\fft{L}{\sqrt{2\left(\ell+\fft{q-1}2\right)}}\left(a_{a,(\ell,m_j)}+a_{a,(\ell,-m_j)}^\dagger\right),\nn\\
    p_{a,(\ell,m_j)}&=\fft1{iL}\sqrt{\fft12\left(\ell+\fft{q-1}2\right)}\left(a_{a,(\ell,-m_j)}-a_{a,(\ell,m_j)}^\dagger\right).
\end{align}
We then find
\begin{equation}
    H=\fft1L\left(N+\sum_{\ell,m_j}\left(\ell+\fft{q-1}2\right)\sum_{a=1}^{p+3}\left(N_{a,(\ell,m_j)}+\fft12\right)+\fft{q+1}2R_1\right),
\end{equation}
where
\begin{align}
    N_{a,(\ell,m_j)}=a_{a,(\ell,m_j)}^\dagger a_{a,(\ell,m_j)}.
\end{align}
Note that, in this basis, the $R_1$ charge operator is 
\begin{equation}
    R_1=-i\sum_{\ell,m_j}\left(a_{p+2,(\ell,m_j)}^\dagger a_{p+3,(\ell,m_j)}-a_{p+3,(\ell,m_j)}^\dagger a_{p+2,(\ell,m_j)}\right).
\end{equation}
The treatment of the angular momenta and charge operators depends on the parity of $p$. For odd $p$, we transform to a helicity basis by taking
\begin{equation}
    a_{i\pm,(\ell,m_j)}=\fft1{\sqrt2}\left(a_{2i-1,(\ell,m_j)}\pm i\,a_{2i,(\ell,m_j)}\right),\qquad i=1,\ldots,(p+3)/2
\end{equation}
For later convenience, we also introduce terminology for fluctuations in this basis:
\begin{align}
    \zeta_{i\pm,(\ell,m_j)} = \dfrac{1}{\sqrt{2}} (\zeta_{2i-1,(\ell,m_j)} \pm i\,\zeta_{2i,(\ell,m_j)}),\qquad i=1,\ldots,(p+3)/2.
\end{align}
The angular momentum and R-charge operators are then
\begin{align}\label{eq:JR_p_odd}
    J_i&=\sum_{\ell,m_j}\left(N_{i+,(\ell,m_j)}-N_{i-,(\ell,m_j)}\right),\qquad i=1,\ldots,(p+1)/2,\nn\\
    R_i&=\begin{cases}
    \begin{array}{rll}
         &\sum_{\ell,m_j}\left(N_{(p+3)/2+,(\ell,m_j)}-N_{(p+3)/2-,(\ell,m_j)}\right),\qquad &i=1,\\
        &\sum_{\ell,m_j}m_{i-1}\sum_{k=1}^{(p+3)/2}\left(N_{k+,(\ell,m_j)}+N_{k-,(\ell,m_j)}\right)\qquad &i=2,\ldots,q,
    \end{array}
    \end{cases}
\end{align}
and the Hamiltonian is
\begin{equation}\label{eq:H_p_odd}
    H=\fft1L\left(N+\sum_{\ell,m_j}\left(\ell+\fft{q-1}2\right)\sum_{i=1}^{(p+3)/2}\left(N_{i+,(\ell,m_j)}+N_{i-,(\ell,m_j)}+1\right)+\fft{q+1}2R_1\right).
\end{equation}
For even $p$, we leave $a_1$ as is and only rotate the other oscillators. Then
\begin{align}
    &a_{1,(\ell,m_j)}=a_{1,(\ell,m_j)},\nn\\
    &a_{i\pm,(\ell,m_j)}=\fft1{\sqrt2}\left(a_{2i-2,(\ell,m_j)}\pm ia_{2i-1,(\ell,m_j)}\right),\qquad i=2,\ldots,(p+4)/2,
\end{align}
with fluctuations
\begin{align}
    \zeta_{i\pm,(\ell,m_j)} = \dfrac{1}{\sqrt{2}} (\zeta_{2i-2,(\ell,m_j)} \pm i\,\zeta_{2i-1,(\ell,m_j)}),\qquad i=2,\ldots,(p+4)/2.
\end{align}
The angular momentum and charge operators are
\begin{align}
    J_i&=\begin{cases}
    \begin{array}{rll}
    &\sum_{\ell,m_j}N_{1,(\ell,m_j)},\qquad&i=1,\\
    &\sum_{\ell,m_j}\left(N_{i+,(\ell,m_j)}-N_{i-,(\ell,m_j)}\right),\qquad &i=2,\ldots,(p+2)/2,
    \end{array}
    \end{cases}\nn\\
    R_i&=\begin{cases}
    \begin{array}{rll}
         &\sum_{\ell,m_j}\left(N_{(p+4)/2+,(\ell,m_j)}-N_{(p+4)/2-,(\ell,m_j)}\right),\qquad &i=1,\\
        &\sum_{\ell,m_j}m_{i-1}\sum_{k=1}^{(p+4)/2}\left(N_{k+,(\ell,m_j)}+N_{k-,(\ell,m_j)}\right)\qquad &i=2,\ldots,q,
    \end{array}
    \end{cases}
\end{align}
and the Hamiltonian is
\begin{equation}
    H=\fft1L\left(N+\sum_{\ell,m_j}\left(\ell+\fft{q-1}2\right)\left(N_{1,(\ell,m_j)}+\sum_{i=2}^{(p+4)/2}\!\!\sum_\pm N_{i\pm,(\ell,m_j)}+\fft{p+3}{2}\right)+\fft{q+1}2R_1\right).
\label{eq:scalarHeven}
\end{equation}
For the evaluation of the scalar index as well as the treatment of fermions and world-volume fields, we consider the D3, M2, and M5 branes separately.

%%%%%%%%%%%%%%%%%%%%%%%%%%%%%%%%%%%%%%%%%%
\section{D3 giant graviton in \texorpdfstring{AdS$_5\times S^5$}{AdS5xS5}}\label{Sec:D3GG}

As the canonical example of AdS/CFT, we start with the case of $\mathcal N=4$ SYM with gauge group $U(N)$, which is dual to AdS$_5\times S^5$.  In this case, the giant graviton expansion of the $\frac{1}{16}$-BPS index involves D3-branes wrapped on an $S^3$ inside $S^5$.  The $\mathcal N=4$ superconformal algebra $SU(2,2|4)$ has the bosonic subalgebra $SO(2,4)\times SO(6)$, and we identify the six Cartan generators as
\begin{equation}
    H,\quad J_1,\quad J_2,\quad R_1,\quad R_2,\quad R_3.
\end{equation}
The $\frac{1}{16}$-BPS index is then given by
\begin{equation}
    \mathcal I_N(p,q;y_i)=\Tr\left((-1)^Fe^{-\beta\mathcal H}p^{J_1}q^{J_2}y_1^{R_1}y_2^{R_2}y_3^{R_3}\right),\qquad pq=y_1y_2y_3,
\end{equation}
where
\begin{equation}
    \mathcal H=H-J_1-J_2-R_1-R_2-R_3.
\end{equation}
The giant graviton expansion takes the form \cite{Imamura:2021ytr,Gaiotto:2021xce}
\begin{equation}
    \fft{\mathcal I_N(p,q;y_i)}{\mathcal I_\infty(p,q;y_i)}=\sum_{m_1,m_2,m_3\ge0}y_1^{m_1N}y_2^{m_2N}y_3^{m_3N}\hat I^{\mathrm{GG}}_{(m_1,m_2,m_3)}(p,q;y_i),
\label{eq:D3gge}
\end{equation}
where the $\{m_i\}$ correspond to the number of wrapped D3-branes moving along the three orthogonal rotation planes corresponding to the $R$-charges $\{R_i\}$.  In the above expansion, the $y_i^{m_iN}$ terms correspond to classical angular momentum on $S^5$ while $\hat I^{\mathrm{GG}}_{(m_1,m_2,m_3)}$ counts the fluctuations of the D3-brane.

Since the giant gravitons are identified as D3-branes, the index for a single stack of $m_1$ giant gravitons is simply the $U(m_1)$ SYM index with fugacities identified as
\begin{equation}
    \hat I^{\mathrm{GG}}_{(m_1,0,0)}(p,q;y_i)=\mathcal I_{m_1}(y_2,y_3;y_1^{-1},p,q).
\end{equation}
The other contributions $\hat I^{\mathrm{GG}}_{(0,m_2,0)}$ and $\hat I^{\mathrm{GG}}_{(0,0,m_3)}$ are obtained by permuting the $\{y_i\}$ fugacities.  Note, however, that the expression for the index is more complicated when there are multiple giants on different rotation planes, as in this case additional states arise at the intersections of the multiple D3-brane giants.

For a single giant graviton, its index is just that of the Abelian theory.  Here the $\mathcal N=4$ SYM index is the plethystic exponential of the single-letter index
\begin{equation}
    f_{\mathrm{SYM}}(p,q;y_i)=1-\fft{(1-y_1)(1-y_2)(1-y_3)}{(1-p)(1-q)}.
\end{equation}
To obtain the $(1,0,0)$ giant graviton index, we map the fugacities
\begin{equation}
    p\to y_2,\quad q\to y_3,\quad y_1\to y_1^{-1},\quad y_2\to p,\quad y_3\to q.
\end{equation}
The single-letter giant graviton index then takes the form 
\begin{equation}
    f_{(1,0,0)}(p,q;y_i)=1-\fft{(1-y_1^{-1})(1-p)(1-q)}{(1-y_2)(1-y_3)}.
\end{equation}
We now demonstrate how this is precisely obtained from the world-volume fluctuations of a single D3-brane giant graviton.

%%%%%%%%%%%%%%%%%
\subsection{Scalar fluctuations}

The world-volume fields of a single D3-brane corresponds to six scalars, one vector and four Majorana fermions.  Following the general analysis of scalar fluctuations above, we consider a D3-brane evolving along time in AdS$_5$ and wrapped on $S^3\subset S^5$.  The six scalars then describe the transverse fluctuations along the four space directions of AdS$_5$ and two directions in $S^5$ orthogonal to the $S^3$.  The flucatuations in AdS$_5$ are quantized with number operators $N_{1\pm}$ and $N_{2\pm}$, while the flucatuations in $S^5$ are quantized with $N_{3\pm}$.  From \eqref{eq:JR_p_odd} and \eqref{eq:H_p_odd} we have
\begin{align}\label{eq:D3_JR}
    J_1&=\sum_{\ell,m_1,m_2}\left(N_{1+,(\ell,m_1,m_2)}-N_{1-,(\ell,m_1,m_2)}\right),\nn\\
    J_2&=\sum_{\ell,m_1,m_2}\left(N_{2+,(\ell,m_1,m_2)}-N_{2-,(\ell,m_1,m_2)}\right),\nn\\
    R_1&=\sum_{\ell,m_1,m_2}\left(N_{3+,(\ell,m_1,m_2)}-N_{3-,(\ell,m_1,m_2)}\right),\nn\\
    R_2&=\sum_{\ell,m_1,m_2}m_1\sum_{i=1}^3\left(N_{i+,(\ell,m_1,m_2)}+N_{i-,(\ell,m_1,m_2)}\right),\nn\\
    R_3&=\sum_{\ell,m_1,m_2}m_2\sum_{i=1}^3\left(N_{i+,(\ell,m_1,m_2)}+N_{i-,(\ell,m_1,m_2)}\right)\nn\\
    H &= \dfrac{1}{L}\Biggl(N + \sum_{\ell.m_1,m_2} (\ell + 1) \sum_{i=1}^2 N_{i+,(\ell,m_1,m_2)} + N_{i-,(\ell,m_1,m_2)}\nn\\
    & \kern4em+ (\ell + 3)N_{3+,(\ell,m_1,m_2)}+ (\ell - 1)N_{3-,(\ell,m_1,m_2)} + 3(\ell + 1)\Biggr),
\end{align}
and we take the supersymmetric Hamiltonian
\begin{align}\label{eq:D3_susyH}
    \mathcal H&=(H\tilde{L}-N)-J_1-J_2-R_1-R_2-R_3\nn\\
    &=\sum_{\ell,m_1m_2}\biggl((\ell-m_1-m_2)\left(N_{1+,(\ell,m_1,m_2)}+N_{2+,(\ell,m_1,m_2)}+N_{3-,(\ell,m_1,m_2)}\right)\nn\\
    &\kern4em+(\ell-m_1-m_2+2)\left(N_{1-,(\ell,m_1,m_2)}+N_{2-,(\ell,m_1,m_2)}+N_{3+,(\ell,m_1,m_2)}\right)+3(\ell+1)\biggr).
\end{align}
Note that we have scaled the hamiltonian by the AdS radius $\tilde{L}$ and also subtracted the classical contribution to the $R_1$ charge, $R_1=N$.  [This is the term responsible for $y_1^{N}$ in the giant graviton expansion (\ref{eq:D3gge})].  Since the scalar $S^3$ harmonics satisfy the condition $|m_1\pm m_2|\le\ell$, we see that the first line can give rise to BPS ($\mathcal H=0$) excitations, provided $m_1+m_2=\ell$.  Excitations from the second line are never BPS, and the final term is the harmonic oscillator zero-point energy, which we discard. We now consider the single-letter scalar partition function
\begin{equation}
    f_{\mathrm{scalar}}=\Tr\left(e^{-\beta\mathcal H}p^{J_1}q^{J_2}y_1^{R_1}y_2^{R_2}y_3^{R_3}\right)=\Tr\left(e^{-\beta\mathcal H}p^{J_1}q^{J_2}y_1^{R_1}(y_2y_3)^{R_L}(y_2/y_3)^{R_R}\right),
\end{equation}
where the trace is over single oscillator excitations, and we have defined the $SO(4)\simeq SU(2)_L\times SU(2)_R$ charges $R_L=\fft12(R_1+R_2)$ and $R_R=\fft12(R_1-R_2)$.  To evaluate the trace, we compute each term arising from the separate oscillators. For each oscillator, the exponents of the fugacities in the index are given by its coefficients in \eqref{eq:D3_JR} and \eqref{eq:D3_susyH}. For example, the contribution from the $N_{1+,(\ell,m_1,m_2)}$ oscillator is
\begin{equation}
    e^{-\beta\mathcal H}p^{J_1}q^{J_2}y_1^{R_1}y_2^{R_2}y_3^{R_3}~\to~ e^{-\beta(\ell-m_1-m_2)}p\,y_2^{m_1}y_3^{m_2}=p\,z^{\ell}(y_2 y_3/z^2)^{m_L}(y_2/y_3)^{m_R}.
\end{equation}
Here we have defined $z=e^{-\beta}$ and replaced $m_1$ and $m_2$ by the $SU(2)_L\times SU(2)_R$ weights $m_L=(m_1+m_2)/2$ and $m_R=(m_1-m_2)/2$.  Since the $SO(4)$ scalar harmonics transform in the $(\ell/2,\ell/2)$ representation, the sum of $m_L$ and $m_R$ from $-\ell$ to $\ell$ yields
an $SO(4)$ character decomposed as two $SU(2)$ characters.  Adding the contributions of all six oscillators, we then find
\begin{equation}
    f_{\mathrm{scalar}}=\left(p+q+y_1^{-1}+z^2(p^{-1}+q^{-1}+y_1)\right)\sum_{\ell\ge0}z^\ell\chi_{\ell/2}(y_2y_3/z^2)\chi_{\ell/2}(y_2/y_3),
\label{eq:D3scalar}
\end{equation}
where $\chi_\ell(x)$ is the $SU(2)$ character
\begin{equation}
    \chi_\ell(x)=\fft{x^{\ell+\fft12}-x^{-\ell-\fft12}}{x^{\fft12}-x^{-\fft12}}.
\end{equation}
Finally, we can perform the sum over $\ell$ to obtain
\begin{equation}
    f_{\mathrm{scalar}}=\fft{(p+q+y_1^{-1}+z^2(p^{-1}+q^{-1}+y_1))(1-z^2)}{(1-y_2)(1-y_3)(1-y_2^{-1}z^2)(1-y_3^{-1}z^2)}.
\label{eq:D3fscalar}
\end{equation}
This is the single-letter scalar partition function of the free theory that includes both BPS and non-BPS states.  The BPS states are those for which $z=0$
\begin{equation}
    f_{\mathrm{scalar}}\Big|_{z=0}=\fft{p+q+y_1^{-1}}{(1-y_2)(1-y_3)}.
\end{equation}
The $z=0$ contribution can also be obtained more directly as follows. Recall that only the $N_{1+}$, $N_{2+}$, and $N_{3-}$ oscillators are BPS and contribute $p\,y_2^{m_1}y_3^{m_2}$, $q\,y_2^{m_1}y_3^{m_2}$, and $y_1^{-1}y_2^{m_1}y_3^{m_2}$ to the index, respectively. To deal with the spherical harmonics, we sum over all BPS states in each symmetric tensor representation of $SO(4)$.  The weight diagrams are diamonds in $[m_1,m_2]$ space whose edges defined by $\pm m_1\pm m_2=\ell$. The BPS states lie on the edge $m_1+m_2=\ell$ with $m_1,m_2\geq 0$. By further summing over $\ell$ we are in fact summing over all integer pairs $[m_1,m_2]$ in the first quadrant of weight space. The sum over scalar BPS states evaluates to
\begin{align}
    f_\text{scalar}\Big|_{\textrm{BPS}}&=(p+q+y_1^{-1})\sum_{\ell=0}^\infty\left(\sum_{\substack{m_1,m_2\geq0\\\sum m=\ell}} y_2^{m_1}y_3^{m_2}\right)=(p+q+y_1^{-1})\!\sum_{m_1,m_2\geq0} y_2^{m_1}y_3^{m_2}\nn\\
    &=\fft{p+q+y_1^{-1}}{(1-y_2)(1-y_3)},
\end{align}
which matches the $z=0$ limit of the scalar partition function computed above.

%%%%%%%%%%%%%%%%%%%%%%
\subsection{Vector fluctuations}

We now turn to the D3-brane world-volume vector field.  Linearizing the DBI action, (\ref{eq:DBI}), results in a standard Maxwell theory
\begin{equation}
    S=\int d^4\xi\sqrt{-g}\left(-\fft14F_{\mu\nu}^2\right),
\end{equation}
where we have canonically normalized the Maxwell field.  Note that this theory is defined on $\mathbb R\times S^3$.  For the free theory, it is straightforward to quantize the abelian gauge field in Coulomb gauge
\begin{equation}
    A_t=0,\qquad\vec\nabla\cdot\vec A=0.
\end{equation}
In this case, the gauge-fixed action is
\begin{equation}
    S=\int d^4\xi\sqrt{-g}\left(\fft12|\dot{\vec A}|^2+\fft12 A_i(\gamma^{ij}\nabla^2-R^{ij})A_j\right),
\end{equation}
where we have integrated by parts.  For the unit $S^3$, we have $R_{ij}=2\gamma_{ij}$.  Furthermore, $S^3$ admits three independent vector spherical harmonics (one longitudinal and two transverse)
\begin{align}
    \vec\nabla^2 \vec X^{(0)}_{\ell,m_1,m_2}&=(2-\ell(\ell+2))\vec X^{(0)}_{\ell,m_1,m_2},&&\vec\nabla\cdot\vec X^{(0)}_{\ell,m_1,m_2}=-\sqrt{\ell(\ell+2)}\vec X^{(0)}_{\ell,m_1,m_2},\nn\\
    \vec\nabla^2 \vec X^{(1)}_{\ell,m_1,m_2}&=(1-\ell(\ell+2))\vec X^{(1)}_{\ell,m_1,m_2},&&\vec\nabla\cdot\vec X^{(1)}_{\ell,m_1,m_2}=0,\nn\\
    \vec\nabla^2 \vec X^{(2)}_{\ell,m_1,m_2}&=(1-\ell(\ell+2))\vec X^{(2)}_{\ell,m_1,m_2},&&\vec\nabla\cdot\vec X^{(2)}_{\ell,m_1,m_2}=0\nn\\
\end{align}
Here $\ell\ge1$, and the two transverse harmonics transform in the $(\fft12(\ell+1),\fft12(\ell-1))$ and $(\fft12(\ell-1),\fft12(\ell+1))$ representations of $SU(2)_L\times SU(2)_R$, respectively, while the longitudinal harmonic transforms as $(\ell/2,\ell/2)$.

Since we work in Coulomb gauge, we only expand in the two transverse harmonics
\begin{equation}
    \vec A=\sum_{\ell,m_1,m_2}A^{(n)}_{(\ell,m_1,m_2)}(t)\vec X^{(n)}_{(\ell,m_1,m_2)}(\xi^i).
\end{equation}
Then
\begin{equation}
    S=\int dt\sum_{\ell,m_1,m_2}\sum_{n={1,2}}\left(\fft12\dot A^{(n)}_{(\ell,m_1,m_2)}\dot A^{(n)}_{(\ell,-m_1,-m_2)}-\fft12(\ell+1)^2A^{(n)}_{(\ell,m_1,m_2)}A^{(n)}_{(\ell,-m_1,-m_2)}\right).
\end{equation}
Note that the range of summation for $m_1$ and $m_2$ are different for the two different transverse vector harmonics.

Quantization is straightforward and proceeds as in the scalar case above.  We find the Hamiltonian
\begin{equation}
    H=\fft1L\sum_{\ell,m_1,m_2}(\ell+1)\left(n^{(1)}_{(\ell,m_1m_2)}+n^{(2)}_{(\ell,m_1,m_2)}+1\right),
\end{equation}
where $n^{(n)}_{(\ell,m_1,m_2)}$ are bosonic number operators.  Here we have restored the length dimension $L$ to the Hamiltonian.  Associated with the gauge excitations, we have
\begin{align}
    J_1&=0,\qquad J_2=0,\qquad R_1=0,\nn\\
    R_2&=\sum_{\ell,m_1,m_2}m_1\left(n^{(1)}_{(\ell,m_1,m_2)}+n^{(2)}_{(\ell,m_1,m_2)}\right),\nn\\
    R_3&=\sum_{\ell,m_1,m_2}m_2\left(n^{(1)}_{(\ell,m_1,m_2)}+n^{(2)}_{(\ell,m_1,m_2)}\right).
\end{align}
The supersymmetric Hamiltonian in the vector sector is then
\begin{align}
    \mathcal H&=\sum_{\ell,m_1,m_2}\left((\ell+1-m_1-m_2)\left(n^{(1)}_{(\ell,m_1m_2)}+n^{(2)}_{(\ell,m_1,m_2)})\right)+(\ell+1)\right)\nn\\
    &=\sum_{\ell,m_1,m_2}\left((\ell+1-2m_L)\left(n^{(1)}_{(\ell,m_1m_2)}+n^{(2)}_{(\ell,m_1,m_2)})\right)+(\ell+1)\right),
\end{align}
where we recall that the sum is over $\ell\ge1$ and where $m_L=(m_1+m_2)/2$ is the $SU(2)_L$ weight.

We now see that zero energy excitations only occur when $m_L=(\ell+1)/2$, which corresponds to maximum $SU(2)_L$ weight in the $(\fft12(\ell+1),\fft12(\ell-1))$ representation corresponding to $n^{(1)}_{(\ell,m_1m_2)}$ excitations.  The contribution to the single-letter partition function is
\begin{equation}
    f_{\mathrm{vector}}=\sum_{\ell\ge1}z^{\ell+1}\left(\chi_{(\ell+1)/2}(y_2y_3/z^2)\chi_{(\ell-1)/2}(y_2/y_3)+\chi_{(\ell-1)/2}(y_2y_3/z^2)\chi_{(\ell+1)/2}(y_2/y_3)\right),
\end{equation}
which sums to
\begin{equation}
    f_{\mathrm{vector}}=\fft{y_2y_3+z^2(y_2+y_3)(y_2^{-1}+y_3^{-1}-2)+z^4(2-2y_2^{-1}-2y_3^{-1}+(y_2y_3)^{-1})}{(1-y_2)(1-y_3)(1-y_2^{-1}z^2)(1-y_3^{-1}z^2)}.
\label{eq:D3vector}
\end{equation}
The sum over BPS states only corresponds to the $z=0$ limit of the partition function
\begin{equation}
    f_{\mathrm{vector}}\Big|_{z=0}=\fft{y_2y_3}{(1-y_2)(1-y_3)}.
\end{equation}
%

%%%%%%%%%%%%%%%%%%%%
\subsection{Fermion fluctuations}

For the fermion sector, we take the fermionic action from \cite{Martucci:2003gc,Marolf:2003ye,Marolf:2003vf,Marolf:2004jb}.  For the D3-brane we have
\begin{equation}
    S=\fft{iT}2\int d^4\xi e^{-\phi}\sqrt{-(g+F)}\bar\theta(1-\tilde\Gamma_{D3})(\gamma^i\hat D_i-\hat\Delta+L_3)\theta,
\end{equation}
where $\theta = (\theta_1,\theta_2)$ is a pair of IIB spinors.  Since we are only interested in fluctuations about the classical background, we have $F_{\mu\nu}=0$ for the world-volume gauge field.  The ten-dimensional geometry is given by the ten-dimensional metric and five-form field strength only.  In this case, we have $L_3=0$ and $\hat\Delta=0$.  The Dirac operator is
\begin{equation}
    \hat D_m=\nabla_m+\fft18\fft1{2\cdot5!}F_5\cdot\Gamma\Gamma_m(i\sigma^2),
\end{equation}
while
\begin{equation}
    \tilde\Gamma_{D3}=\fft{-i\sigma^2}{\sqrt{-\gamma}}\fft1{4!}\epsilon^{ijkl}\Gamma_{ijkl}=\Gamma_{0123}(-i\sigma^2),
\end{equation}
where $0,1,2,3$ denote the D3-brane world-volume tangent space indices and the Pauli matrix $i\sigma^2$ acts on the IIB spinor pair.  Using the background (\ref{eq:AdSS}) and its natural vielbein basis, we have
\begin{equation}
    S=iT\int dt d^3\xi L^3\sqrt\gamma\,\bar\theta\,\fft{1-\tilde\Gamma_{D3}}2\,\Gamma^i\left(\nabla_i+\fft18\fft1{2\cdot5!}F_5\cdot\Gamma\Gamma_i(i\sigma^2)\right)\theta,
\end{equation}
with
\begin{equation}
    \tilde\Gamma_{D3}=\Gamma^{0123}(i\sigma^2).
\end{equation}
We take the convention that $0,1,2,3$ correspond to $t$ and the coordinates $\xi^1,\xi^2,\xi^3$ on $S^3$, while $4,5,6,7$ correspond to the space coordinates $\zeta_a$ on AdS$_5$ and $8,9$ correspond to $x,y$ on $S^5$.  In this case
\begin{equation}
    \fft1{5!}F_5\cdot\Gamma=-\fft4L\Gamma^{12389}(1+\Gamma^{11})\qquad\Rightarrow\qquad\fft1{5!}\Gamma^iF_5\cdot\Gamma\Gamma_i=-\fft8L\Gamma^{12389}(1-\Gamma^{11}).
\label{eq:F5dotGamma}
\end{equation}
The $\kappa$ symmetry projects onto spinors $\theta$ with $\tilde\Gamma_{D3}=1$.  This allows us to identify $(i\sigma^2)=-\Gamma^{0123}$, which relates $\theta_1$ with $\theta_2$.  Furthermore, from (\ref{eq:F5dotGamma}), we see that in these conventions the ten-dimensional Weyl condition is $\Gamma^{11}\theta=-\theta$. This reduces the fermionic action to,
\begin{equation}
    S=iT\int dtd^3\xi L^3\sqrt h\bar\theta_1\left(\Gamma^0\partial_t+\fft1L\fsl\nabla+\fft1L\Gamma^{089}\right)\theta_1,
\end{equation}
We can now decompose the ten-dimensional Majorana-Weyl spinor, $\theta_1$, into a set of four dimensional Weyl-spinors $\psi$ on the world-volume, which transforms according to the $4$ and $\bar 4$ of the transverse $SO(6)$. Since $\Gamma^{11}\theta_1 = -\theta_1$ and $B_{10} \theta_1 = \theta_1^*$, following the conventions in Appendix~\ref{app: Spinors 10 D} our decomposition must be of the form,
\begin{align}
    \theta_1 = \sum_p
    \begin{pmatrix}
        1\\
        0
    \end{pmatrix}
    \otimes
    \psi_p \otimes \eta^{(+)}_p + 
    \begin{pmatrix}
        0\\
        1
    \end{pmatrix}
    \otimes
    \sigma^2\psi^*_p \otimes \tilde b_6 \eta^{(+)}_p\nn,
\end{align}
where $p$ runs over all the states in $\eta^{(\pm)}$ and where we have further decomposed the $SO(1,3)$ Dirac matrices according to (\ref{eq:Dirac3+1}).
This gives
\begin{equation}
    S=\dfrac{iT}{L} \sum_p \int dtd^3\xi L^3\sqrt h\psi^\dagger_p\left(-\partial_t+\fsl\nabla+2i R_1\right)\psi_p + c.c,
\end{equation}
where $R_1=\fft{i}2\Gamma^{89}$. Similar to the bosonic case, we decompose $\psi$ into spinor harmonics on $S^3$
\begin{equation}
    \psi = \sum_{s,n,m_1,m_2} \psi^s_{(n,m_1,m_2)}\chi^s_{(n,m_1,m_2)}, \qquad s = \pm,
\end{equation}
where
\begin{equation}
    \fsl\nabla\chi^s_{(n,m_1,m_2)}= is \left(n+\fft32\right)\chi^s_{(n,m_1,m_2)},\quad n=0,1,2,\ldots.
\end{equation}
These spinor harmonics transform as $(n/2,(n+1)/2)$ and $((n+1)/2,n/2)$ under $SU(2)_L\times SU(2)_R$.  The action then takes the form
\begin{equation}
    S=\fft{N}L\int dt\sum_{p,s,n,m_1,m_2}\psi^s{}^\dagger_{p,(n,m_1,m_2)}\left(-i\partial_t-s\left(n+\fft32\right)-2R_1\right)\psi^s_{p,(n,m_1,m_2)}+.
\end{equation}
The Hamiltonian then takes the form,
\begin{equation}
    H = \dfrac{1}{L} \sum_{p,s,n,m_1,m_2} \left(s \left(n+\dfrac{3}{2}\right)+2R_1\right) \psi^s{}^\dagger_{p,(n,m_1,m_2)} \psi^s_{p,(n,m_1,m_2)}.
\end{equation}
For supersymmetry ($\kappa$-symmetry) the decomposition should be according to the quantum numbers
\begin{center}
\begin{tabular}{ccc|c}
$J_1$&$J_2$&$R_1$&$H$\\
\hline
$+$&$+$&$-$&$n+1/2$\\
$-$&$-$&$-$&$n+1/2$\\
$+$&$-$&$+$&$n+5/2$\\
$-$&$+$&$+$&$n+5/2$\\
\hline
$+$&$-$&$-$&$n+1/2$\\
$-$&$+$&$-$&$n+1/2$\\
$+$&$+$&$+$&$n+5/2$\\
$-$&$-$&$+$&$n+5/2$\\\end{tabular}
\end{center}
The first group corresponds to $\bar4$ of transverse $SO(6)$ transforming as $(n/2,(n+1)/2)$ of $SU(2)_L\times SU(2)_R$, while the second group corresponds to $4$ and  $((n+1)/2,n/2)$ of $SU(2)_L\times SU(2)_R$. 
Then the contribution to the single-letter fermion partition function takes the form
\begin{align}
    f_{\mathrm{fermion}}&=\sum_{n\ge0}z^n\left[\left(\fft{pq}{y_1}\right)^{\fft12}+z^2\left(\left(\fft1{pqy_1}\right)^{\fft12}+\left(\fft{py_1}q\right)^{\fft12}+\left(\fft{qy_1}p\right)^{\fft12}\right)\right]\chi_{n/2}(y_2y_3/z^2)\chi_{(n+1)/2}(y_2/y_3)\nn\\
    &\kern2em+z^n\left[z^3\left(\fft{y_1}{pq}\right)^{\fft12}+z\left(\left(pqy_1\right)^{\fft12}+\left(\fft{q}{py_1}\right)^{\fft12}+\left(\fft{p}{qy_1}\right)^{\fft12}\right)\right]\chi_{(n+1)/2}(y_2y_3/z^2)\chi_{n/2}(y_2/y_3).
\end{align}
Performing the sum over $n$ gives
\begin{align}
    f_{\mathrm{fermion}}&=-\fft1{(pqy_1y_2y_3)^{\fft12}}\fft{(y_2y_3-y_2-y_3+z^2)(pq+z^2(1+y_1(p+q)))}{(1-y_2)(1-y_3)(1-z^2/y_2)(1-z^2/y_3)}\nn\\
    &\quad+\fft1{(pqy_1y_2y_3)^{\fft12}}\fft{(p+q+pqy_1+z^2y_1))(y_2y_3+z^2(1-y_2-y_3))}{(1-y_2)(1-y_3)(1-z^2/y_2)(1-z^2/y_3)},
\label{eq:D3fermion}
\end{align}
while the sum over BPS states is
\begin{equation}
    f_{\mathrm{fermion}}\Big|_{z=0}=\fft{y_2+y_3-y_2 y_3+pq+p/y_1+q/y_1}{(1-y_2)(1-y_3)}.
\end{equation}
%

%%%%%%%%%%%%%
\section{ Indices from excitations}\label{Sec:PartFn}

The single-letter D3-brane giant graviton index is obtained by adding together the scalar, (\ref{eq:D3scalar}), vector, (\ref{eq:D3vector}), and fermion, (\ref{eq:D3fermion}), contributions
\begin{equation}\label{eq:f(1,0,0)}
    f_{(1,0,0)}=f_{\mathrm{scalar}}+f_{\mathrm{vector}}-f_{\mathrm{fermion}}=1-\fft{(1-y_1^{-1})(1-p)(1-q)}{(1-y_2)(1-y_3)}.
\end{equation}
Note that the $z$ dependence of the individual partition functions cancel, as expected for the index. This result exactly matches those of \cite{Imamura:2021ytr,Gaiotto:2021xce}. The giant graviton indices $f_{(0,1,0)}$ and $f_{(0,0,1)}$ are obtained from \eqref{eq:f(1,0,0)} via permutation of the $y$ fugacities.

The full list of D3 brane modes, both BPS and non-BPS, is presented in Table \ref{tab: D3 fluctuations}. Though not explicitly shown in the table, the fluctuations may be further organized into multiplets under the $\mathrm{SO}(4)_J$ symmetry of AdS$_5$. The scalars $\zeta_{1\pm}$, $\zeta_{2\pm}$ transform in the $\mathbf{4}$, and the fermions combine into four doublets $\bar\psi_{\pm\pm-}$, $\bar\psi_{\pm\mp+}$, $\psi_{\pm\mp-}$, and $\psi_{\pm\pm+}$. The remaining modes are singlets. The $\mathrm{SO}(4)_J$ multiplets precisely correspond to the entries in Table 2 of \cite{Imamura:2021ytr}, up to a different definition of the parameter $\ell$.

\begin{table}[ht]
    \centering
    \begin{tabular}{c c c c c c c c}
        \toprule
        {}  & $H\tilde L$ & $J_1$ & $J_2$ & $R_1$& $[R_2,R_3]$& $\mathcal{H}$  \\
        \midrule
        $\zeta_{1+,(\ell,m_1,m_2)}$& $\ell + 1$ & $1$ & $0$ & $0$ & $[\ell,0]$ & $\ell - (m_1 + m_2)$ \\
        $\zeta_{2+,(\ell,m_1,m_2)}$& $\ell + 1$ & $0$ & $1$ & $0$ & $[\ell,0]$ & $\ell - (m_1 + m_2)$\\
        $\zeta_{3+,(\ell,m_1,m_2)}$& $\ell + 3$ & $0$ & $0$ & $1$ & $[\ell,0]$ & $\ell + 2 - (m_1 + m_2)$ \\ 
        $\zeta_{1-,(\ell,m_1,m_2)}$& $\ell + 1$ & $-1$ & $0$ & $0$ & $[\ell,0]$ & $\ell + 2 - (m_1 + m_2)$\\
        $\zeta_{2-,(\ell,m_1,m_2)}$& $\ell + 1$ & $0$ & $-1$ & $0$ & $[\ell,0]$ & $\ell +2 - (m_1 + m_2)$\\
        $\zeta_{3-,(\ell,m_1,m_2)}$& $\ell -1$ & $0$ & $0$ & $-1$ & $[\ell,0]$ & $\ell - (m_1 + m_2)$\\ 
        \midrule
        $A^{(1)}_{\ell,m_1,m_2}$ & $\ell+1$ & $0$ & $0$ & $0$ & $[\ell,1]$ & $\ell+1 - (m_1 + m_2)$\\
        $A^{(2)}_{\ell,m_1,m_2}$ & $\ell+1$ & $0$ & $0$ & $0$ & $[\ell,-1]$ & $\ell+1 - (m_1 + m_2)$\\
        \midrule
        $\bar\psi_{++-,(\ell,m_1,m_2)}$ & $\ell + 1/2$ & $1/2$ & $1/2$ & $-1/2$ & $[\ell+1/2,-1/2]$ & $\ell - (m_1 + m_2)$\\
        $\bar\psi_{---,(\ell,m_1,m_2)}$ & $\ell + 1/2$ & $-1/2$ & $-1/2$ & $-1/2$ & $[\ell+1/2,-1/2]$ & $\ell+2 - (m_1 + m_2)$ \\
        $\bar\psi_{+-+,(\ell,m_1,m_2)}$ & $\ell + 5/2$ & $1/2$ & $-1/2$ & $1/2$ & $[\ell+1/2,-1/2]$ & $\ell+2 - (m_1 + m_2)$ \\
        $\bar\psi_{-++,(\ell,m_1,m_2)}$ & $\ell + 5/2$ & $-1/2$ & $1/2$ & $1/2$ & $[\ell+1/2,-1/2]$ & $\ell+2 - (m_1 + m_2)$ \\
        $\psi_{+--,(\ell,m_1,m_2)}$ & $\ell + 1/2$ & $1/2$ & $-1/2$ & $-1/2$ & $[\ell+1/2,1/2]$ & $\ell+1 - (m_1 + m_2)$\\
        $\psi_{-+-,(\ell,m_1,m_2)}$ & $\ell + 1/2$ & $-1/2$ & $1/2$ & $-1/2$ &  $[\ell+1/2,1/2]$ &$\ell+1 - (m_1 + m_2)$\\
        $\psi_{+++,(\ell,m_1,m_2)}$ & $\ell + 5/2$ & $1/2$ & $1/2$ & $1/2$ &  $[\ell+1/2,1/2]$ & $\ell + 1 - (m_1 + m_2)$\\
        $\psi_{--+,(\ell,m_1,m_2)}$ & $\ell + 5/2$ & $-1/2$ & $-1/2$ & $1/2$ &  $[\ell+1/2,1/2]$ & $\ell +3 - (m_1 + m_2)$ \\
        \bottomrule
    \end{tabular}
    \caption{Quantum numbers of the $D_3$ brane fluctuation corresponding to each letters.We have subtracted the ground state and classical contribution from $H$ in this table. Here $[R_2, R_3]$ labels the representation of $SO(4)$ with quantum numbers of the highest weight state and $m_1,m_2$ are the quantum numbers of the general state in the representation $[R_2,R_3]$. The $\pm$ subscripts on the spinors corresponds to their $J_1, J_2, R_1$ charges.}
    \label{tab: D3 fluctuations}
\end{table}

\subsection{The \texorpdfstring{$\fft{1}{2}$-BPS}{1/2-BPS} index}\label{sec:1/2BPS} 
One may also take limits on the fugacities in \eqref{eq:f(1,0,0)} to arrive at various giant graviton expansions previously studied in the literature. Consider the $\frac12$-BPS index
\begin{equation}\label{eq:1/2BPS}.
        \mathcal{I}_N(y_1)=\prod_{n=1}^N\frac1{1-y_1^n},
\end{equation}
which counts operators in $\mathcal{N}=4$  SYM consisting of multitraces of a single complex scalar with $H=R_1$. The $\frac12$-BPS index is recovered from the full index by taking the limit $p,q,y_2,y_3\to0$. Applying the same fugacity limit to the giant graviton index, \eqref{eq:f(1,0,0)}, reduces to $f_{(1,0,0)}=y_1^{-1}$. This term is just the contribution of the scalar mode $\zeta_{3-}$ with $l=0$, which carries $R_1$ charge $-1$. Plethystic exponentiation gives
\begin{equation}\label{eq:halfBPS_PE}
        \text{PE}[y_1^{-1}]=\fft1{1-y_1^{-1}},
\end{equation}
which matches the $m=1$ term in the giant graviton expansion of the $\frac12$-BPS index,
\begin{equation}\label{eq:1/2BPS_GGE}
        \mathcal{I}_N(y_1)=\mathcal{I}_\infty(y_1)\sum_{m=0}^\infty y_1^{mN}\prod_{k=1}^m\fft1{1-y_1^{-k}}.
\end{equation}
Later in section \ref{Sec:nonAb}, we will derive the $\frac12$-BPS GGE to all orders by promoting the scalar fluctuation to an $m\times m$ matrix-valued field. 

Some previous studies of the $\frac12$-BPS giant graviton expansion appear in \cite{Lee:2022vig,Lee:2023iil,Eleftheriou:2023jxr,Lee:2024hef,Eleftheriou:2025lac,Chang:2024zqi,Deddo:2024liu}. The negative $R$ charge fluctuations in the GGE have been equivalently interpreted as modifications of the  determinant operators dual to giant gravitons \cite{Lee:2022vig}, as well as fluctuations of fully backreacted supergravity geometries \cite{Deddo:2024liu}.

The $\frac12$-BPS GGE was also previously derived from the D3 brane DBI action in \cite{Lee:2023iil,Eleftheriou:2023jxr,Lee:2024hef}. These approaches instead result in an analytically continued form of \eqref{eq:1/2BPS_GGE} involving only positive powers of the fugacity, but with alternating signs inside the sum. Naturally, this other form of the GGE requires a somewhat different interpretation. In \cite{Eleftheriou:2023jxr} the alternating signs are explained by fermionic states arising from a supersymmetrized fluctuation action, and in \cite{Lee:2023iil} they are interpreted as an extra fermionic grading of the bulk Hilbert space.

\subsection{The Schur index} 

Another limit of the index is the Schur index, originally defined for $\mathcal{N}=2$ theories in \cite{Gadde:2011ik,Gadde:2011uv}. The expansion at large $N$ appears in \cite{Bourdier:2015sga,Bourdier:2015wda}. With our fugacity conventions, the $\mathcal{N}=4$ SYM Schur index is defined by setting $q=y_3$. Since $pq=y_1y_2y_3$, the condition $p=y_1y_2$ is also enforced automatically. 
This identification of fugacities naturally results in extra cancellations of states with different quantum numbers.
Due to these cancellations, the Schur index actually depends on only two fugacities $y_1,y_2$ rather than the naive three \cite{Gadde:2011uv}.
From equation \eqref{eq:f(1,0,0)} and its permutations, we obtain the giant graviton indices
\begin{align}
    f_{(1,0,0)}^\text{Schur}(y_1,y_2)&=1-\fft{(1-y_1^{-1})(1-y_1y_2)}{(1-y_2)}\nn\\
    f_{(0,1,0)}^\text{Schur}(y_1,y_2)&=1-\fft{(1-y_2^{-1})(1-y_1y_2)}{(1-y_1)}\nn\\
    f_{(0,0,1)}^\text{Schur}(y_1,y_2,y_3)&=1-\frac{(1-y_3^{-1})(1-y_1 y_2)(1-y_3)}{(1-y_1)(1-y_2)}.
\end{align}
From the overall $y_3$ independence of the Schur index, one expects that the $(0,0,1)$ giant should not contribute. This can be seen explicitly via a series expansion and a more careful limit $q\to y_3$ :
\begin{equation}
    f_{(0,0,1)}^\text{Schur}(y_1,y_2,y_3)=\lim_{q\to y_3} \left(y_3^{-1}+q-q y_3^{-1}+\mathcal{O}(y_1,y_2)\right).
\end{equation}
The $q/y_3$ term sends the plethystic exponential to zero
\begin{equation}
    \text{PE}\left[f_{(0,0,1)}^\text{Schur}(y_1,y_2,y_3)\right]=\lim_{q\to y_3} \frac1{1-y_3^{-1}}\frac1{1-q}(1-q y_3^{-1})\cdots=0,
\end{equation}
so the $(0,0,1)$ giant does not contribute to the Schur index.

Along with the two-fugacity Schur index, one may also consider a less refined limit with $x\equiv y_1=y_2$:
\begin{equation}
    f_{(1,0,0)}^\text{Schur}(x)=f_{(0,1,0)}^\text{Schur}(x)=1+ x^{-1}-x.
\end{equation}
This result matches \cite{Beccaria:2024vfx,Gautason:2024nru}, which also presented a holographic derivation of the first term in the Schur giant graviton expansion.
Because both single-giant indices are equal, one can compute the leading term from the fluctuations of just one type of giant.

It is also worth explicitly tracking the states that contribute to this answer after cancelations are taken into account. To explicitly discard fluctuations that give no net contribution to the giant graviton expansion, we make use of the $q,y_3$ independence of the Schur index and send $q=y_3\to 0$:
\begin{eqnarray}
    f_{scalar}&=& \fft{p+q+y_1^{-1}}{(1-y_2)(1-y_3)}\;\stackrel{q=y_3=0}{\mapsto}\;\fft{p+y_1^{-1}}{1-y_2}\mapsto\fft{x^2+x^{-1}}{1-x}\nn \\
    f_{\rm vector}&=& \fft{y_2y_3}{(1-y_2)(1-y_3)} \;\stackrel{q=y_3=0}{\mapsto}\;0, \nn \\
    f_{\rm fermions}&=& \fft{y_2+y_3-y_2 y_3+pq+p/y_1+q/y_1}{(1-y_2)(1-y_3)}\;\stackrel{q=y_3=0}{\mapsto}\;\frac{y_2+p/y_1}{1-y_2}\mapsto\fft{2x}{1-x}\nn \\
    f_{(1,0,0)}^\text{Schur}&=&f_{\rm scalar}-f_{\rm fermions}=\fft{x^2+x^{-1}-2x}{1-x}=1+x^{-1}-x.
\end{eqnarray}
By tracing the surviving terms back to Table \ref{tab: D3 fluctuations}, we find that the only net contribution to $f_{(1,0,0)}^\text{Schur}$ comes from two scalar modes $\zeta_{1+}$, $\zeta_{3-}$, two fermionic modes $\bar\psi_{++-}$, $\psi_{+--}$, and some of their spherical harmonics.

%%%%%%%%%%%%%%%%%%%%%%%%%%%%%%%%%%%%%%%%%%%%%%%%%%%%%%
\subsection{The complete superconformal index}

The fluctuation analysis described so far produces the single-giant terms in the GGE. There are two obstructions to obtaining the rest of \eqref{eq:D3gge} in this manner. First, terms of the form $\hat I^{\mathrm{GG}}_{(m,0,0)}$ for arbitrary $m$ correspond to fluctuations of stacks of giants, which are not captured by the DBI action. However,  $\hat I^{\mathrm{GG}}_{(m,0,0)}$  can still be determined without the explicit fluctuations once $f_{(1,0,0)}$ is known \cite{Imamura:2021ytr,Gaiotto:2021xce,Lee:2022vig}. To effectively promote single-giant fluctuations to strings between multiple branes, one appends adjoint characters to the single-letter index. Taking the plethystic exponential and projecting to $\mathrm{U}(m)$ singlet states gives
\begin{equation}
    \hat I^{\mathrm{GG}}_{(m,0,0)}(p,q,y_i)=\frac1{m!}\prod_{a=1}^m\frac{d\sigma_a}{2\pi i \sigma_a}\prod_{a\neq b}\left(1-\frac{\sigma_a}{\sigma_b}\right)\prod_{a=1}^m\prod_{b=1}^m \mathrm{PE}\left[f_{(1,0,0)}(p,q,y_i)\left(\frac{\sigma_a}{\sigma_b}+\frac{\sigma_b}{\sigma_a}\right)\right].
\end{equation}

In Section \ref{Sec:nonAb}, we instead perform an explicit computation of $\hat I^{\mathrm{GG}}_{(m,0,0)}$ for the $\frac12$-BPS index by promoting the scalar fluctuation to a $m\times m$ matrix, as required by non-abelian generalizations of the DBI action. The result matches the integral form above. Although we do not extend the analysis to the full index, such a non-abelian generalization should be possible in principle.

Besides stacks of coincident giants, the remaining part of the giant graviton expansion also involves terms $\hat I^{\mathrm{GG}}_{(m_1,m_2,m_3)}$ with arbitrary $m_1,m_2,m_3$ corresponding to orthogonal stacks of giants. In this case there are also hypermultiplets at the intersections of branes. Although the charges for these excitations were fully determined in \cite{Imamura:2021ytr}, such modes are not accessible with our DBI analysis.
%
%%%%%%%%%%%%%%%%%%%%%%%%%%%%%%%%%%%%%%%%%%%%%%%%
\section{M5 giant graviton in \texorpdfstring{AdS$_4\times S^7$}{AdS4xS7}}\label{Sec:M5}

We now turn to the M-brane indices, starting with the near-horizon limit of M2-branes giving rise to ABJM theory with $k=1$ dual to AdS$_5\times S^7$.  This theory has the $\mathcal{N}=8$ superconformal algebra $OSp(8|4)$ with bosonic subalgebra $SO(2,3)\times SO(8)$.  We denote the six Cartan generators by
\begin{equation}
    H,\quad J,\quad R_1,\quad R_2,\quad R_3,\quad R_4,
\end{equation}
corresponding to the Hamiltonian, spin and four $R$-charges, respectively. The superconformal index may be defined as \cite{Arai:2020uwd,Beccaria:2023cuo}\footnote{Our conventions relate to those of \cite{Arai:2020uwd} by $y_i=q^{1/2}u_i$ and $p=q^2$.}
\begin{equation}
    \mathcal{I}_N(p;y_i)=\operatorname{Tr}\left[(-1)^F e^{-\beta\mathcal H}p^J y_1^{R_1}y_2^{R_2}y_3^{R_3}y_4^{R_4}\right],\qquad p=y_1 y_2 y_3 y_4,
\end{equation}
where
\begin{equation}
    \mathcal H=H-J-\fft12(R_1+R_2+R_3+R_4).
\label{eq: delta AdS4}
\end{equation}

The giant gravitons in the expansion of the ABJM index are M5-branes wrapped on an $S^5$ inside $S^7$.  Since $S^7$ incorporates four orthogonal rotation planes (corresponding to the four $R$-charges), giant graviton configurations are described by four integers $\{m_i\}$ indicating the number of giant gravitons associated with each of the planes.  The giant graviton expansion of the ABJM index then takes the form
\begin{equation}\label{eq:ABJM_gge}
    \fft{\mathcal I_N(p;y_i)}{\mathcal I_\infty(p;y_i)}=\sum_{m_i\ge0}y_1^{m_1N}y_2^{m_2N}y_3^{m_3N}y_4^{m_4N}\hat I^{\mathrm{GG}}_{(m_1,m_2,m_3,m_4)}(p;y_i).
\end{equation}
The single giant graviton indices correspond to $\hat I^{\mathrm{GG}}_{(1,0,0,0)}$ and similar versions for the other rotation planes.  Since a single M5-brane is described by the abelian $(2,0)$ theory, its index is simply given by $\mathcal I_{M5}=\mathrm{PE}(f_{M5})$ where
\begin{equation}
    f_{M5}(p_1,p_2,p_3;y_1,y_2)=\fft{y_1+y_2-y_1y_2(p_1^{-1}+p_2^{-1}+p_3^{-1})+y_1y_2}{(1-p_1)(1-p_2)(1-p_3)},\qquad p_1p_2p_3=y_1y_2.
\end{equation}
Here $p_i$ are fugacities associated with the angular momentum generators $J_i$ and $y_i$ are fugacities associated with the $R$-charge generators $R_i$ under the decomposition of the $(2,0)$ superalgebra $OSp(8^*|4)\supset SO(2,6)\times SO(5)$.  The giant graviton index $\hat I^{\mathrm{GG}}_{(1,0,0,0)}$ is obtained by mapping the $OSp(8^*|4)$ fugacities to the corresponding $OSp(8|4)$ fugacities
\begin{equation}
    p_1\to y_2,\quad p_2\to y_3,\quad p_3\to y_4,\quad y_1\to-y_1,\qquad y_2\to p.
\end{equation}
This results in $\hat I^{\mathrm{GG}}_{(1,0,0,0)}=\mathrm{PE}(f_{(1,0,0,0)})$ where
\begin{align}
    f_{(1,0,0,0)}(p;y_i)&=\fft{y_1^{-1}+p-py_1^{-1}(y_2^{-1}+y_3^{-1}+y_4^{-1})+py_1^{-1}}{(1-y_2)(1-y_3)(1-y_4)}\nn\\
    &=\fft{y_1^{-1}+p-(y_3 y_4+y_2 y_4+ y_2 y_3)+y_2 y_3 y_4}{(1-y_2)(1-y_3)(1-y_4)}.
\label{eq:M5ggindex}
\end{align}
This is the result we shall derive from our M5-brane fluctuation analysis.

%%%%%%%%%%%%%%%%%%%%%%%%%%%%%%%%%%%%%
\subsection{Scalar fluctuations}

The world-volume fields of a single M5 correspond to a $(2,0)$ tensor multiplet with five scalars transforming as the $\mathbf 5$ of $SO(5)\simeq Sp(4)$, an antisymmetric tensor $B_{\mu\nu}$ with anti-self-dual field strength and four symplectic Majorana Weyl fermions transforming as the $\mathbf 4$ of $Sp(4)$.

We start with the scalar fluctuations, where the Hamiltonian is given by (\ref{eq:scalarHeven}) which for $p = 2$ takes the form,
\begin{align}
    H &= \dfrac{1}{2\tilde L} \Biggl( N + \sum_{\ell,m_j} (\ell + 2) (N_{1,(\ell,m_j)}+ N_{2+,(\ell,m_j)} + N_{2-,(\ell,m_j)} \\
    &\kern4em+ (\ell + 5) N_{3+,(\ell,m_j)}+ (\ell - 1) N_{3-,(\ell,m_j)})+ \dfrac{5}{2}(\ell + 2) \Biggr)
\end{align}

The corresponding supersymmetric Hamiltonian, (\ref{eq: delta AdS4}), is then written as
\begin{align}
    \mathcal H&=\left(H\tilde{L}-\fft{N}2\right)-J-\fft12\sum_{i=1}^4 R_i\nn\\
    &=\fft12\sum_{\ell,m_j}\biggl(\left(\ell-\sum m_j\right)\left(N_{2+,(\ell,m_j)}+N_{3-,(\ell,m_j)}\right)+\left(\ell-\sum m_j+2\right)N_{1,(\ell,m_j)}\nn\\
    &\kern4em+\left(\ell-\sum m_j+4\right)\left(N_{2-,(\ell,m_j)}+N_{3+,(\ell,m_j)}\right)+\fft52(\ell+2)\biggr).
\label{eq:calHscalar}
\end{align}
The conserved charges corresponding to the five scalar fluctuations are
\begin{align}
    J&=\sum_{\ell,m_j}(N_{2+,(\ell,m_j)}-N_{2-,(\ell,m_j)}),\nn\\
    R_1&=\sum_{\ell,m_j}(N_{3+,(\ell,m_j)}-N_{3-,(\ell,m_j)}),\nn\\
    R_i&=\sum_{\ell,m_j}m_i(N_{1,(\ell,m_j)}+N_{2+,(\ell,m_j)}+N_{2-,(\ell,m_j)}+N_{3+,(\ell,m_j)}+N_{3-,(\ell,m_j)}).
\label{eq:ccharges}
\end{align}
As a result, the single-letter scalar partition function takes the form
\begin{equation}
    f_{\mathrm{scalar}}=(p+y_1^{-1}+z+z^2(p^{-1}+y_1))\sum_{\ell,m_j}z^{\ell/2}\prod_j(y_j/\sqrt{z})^{m_j},
\end{equation}
where the sum is over the scalar harmonics of $S^5$.

The scalar harmonics transform as the $\ell$-fold symmetrized tensor representation of SO(6).  Labeling SO(6) weights as $[q_1,q_2,q_3]$ under the Cartan generators $R_1$, $R_2$ and $R_3$, respectively, this corresponds to the highest-weight states $[\ell,0,0]$, where $[1,0,0]$ labels the vector of SO(6).  Using the map between SO(6) highest weights and SU(4) Dynkin labels $(a_1,a_2,a_3)$
\begin{equation}
    a_1=q_2+q_3,\qquad a_2=q_1-q_2,\qquad a_3=q_2-q_3,
\label{eq:Dynkin}
\end{equation}
we see that the scalar harmonics transform in the $(0,\ell,0)$ representation of SU(4).  As a result, the scalar partition function is a sum of SU(4) characters
\begin{equation}
    f_{\mathrm{scalar}}=(p+y_1^{-1}+z+z^2(p^{-1}+y_1))\sum_{\ell\ge0}z^{\ell/2}\chi_{(0,\ell,0)}(y_jz^{1/4}/\sqrt{y_2y_3y_4}).
\label{eq:su4fscalar}
\end{equation}
Here we have converted the SO(6) fugacities $y_j/\sqrt{z}$ into SU(4) fugacities
\begin{equation}
    x_1=\fft{y_2z^{1/4}}{\sqrt{y_2y_3y_4}},\qquad x_2=\fft{y_3z^{1/4}}{\sqrt{y_2y_3y_4}},\qquad x_3=\fft{y_4z^{1/4}}{\sqrt{y_2y_3y_4}},\qquad x_4=\sqrt{y_1y_2y_3}z^{-3/4},
\end{equation}
and the SU(4) character in the fundamental is
\begin{equation}
    \chi_{(1,0,0)}(x_i)=x_1+x_2+x_3+x_4.
\end{equation}

The SU(4) characters can be written out explicitly using the Weyl character formula (see Appendix~\ref{app:characters}).   Performing the sum over $\ell$ then gives
\begin{equation}
    f_{\mathrm{scalar}}=\fft{(p+y_1^{-1}+z+z^2(p^{-1}+y_1))(1-z)}{(1-y_2)(1-y_3)(1-y_4)(1-z/y_2)(1-z/y_3)(1-z/y_4)}.
\label{eq:IM5scalar}
\end{equation}
To find the contribution to the index, we can take $z=0$, in which case we obtain
\begin{equation}\label{eq:M5_scalar_index}
    f_{\mathrm{scalar}}\Big|_{z=0}=\fft{p+y_1^{-1}}{(1-y_2)(1-y_3)(1-y_4)}.
\end{equation}
Like the D3 brane scalar index, \eqref{eq:M5_scalar_index} can also be computed more directly by restricting to BPS excitations from the start. Only $N_{2+}$ and $N_{3-}$ are BPS, and these excitations contribute $p\,y_2^{m_1}y_3^{m_2}y_4^{m_3}$ and $y_1^{-1}y_2^{m_1}y_3^{m_2}y_4^{m_3}$ to the index, respectively. The weight diagrams for the $SO(6)$ harmonics are octahedrons in $(m_1,m_2,m_3)$ space, and only one of the faces is BPS. Analogously to D3 brane scalar index, the denominator in \eqref{eq:M5_scalar_index} arises from a sum over the entire first octant of weight space. 
%%%%%%%%%%%%%%%%%%%%%%%%%%%%%%%%%%%%%%%%%%
\subsection{Two-form fluctuations}

In order to quantize the antisymmetric tensor on the world-volume of the M5-brane, which has an anti-self-dual field strength, we start with a covariant action
\begin{equation}
    S=\int d^6\xi\sqrt{-g}\left(-\fft1{12}F_{\mu\nu\rho}F^{\mu\nu\rho}\right), \qquad F_{\mu\nu\rho}=3\nabla_{[\mu}A_{\nu\rho]},
\end{equation}
and impose anti-self-duality on the spectrum after quantization.  As in the Maxwell case, quantization is easily performed in the Coulomb gauge:
\begin{equation}
    A_{j0}=0,\qquad \nabla^iA_{ij}=0 \qquad j=1,\ldots,5.
\end{equation}
In this case, the action becomes
\begin{equation}
    S=\int d^6\xi\sqrt{-g}\left(\fft14(\dot{A}_{ij})^2-\fft1{12}\left(\nabla_iA_{jk}+\nabla_jA_{ki}+\nabla_kA_{ij}\right)^2\right).
\end{equation}
Applying integration by parts to the second term and commuting derivatives gives
\begin{align}
    \left(\nabla_iA_{jk}+\nabla_jA_{ki}+\nabla_kA_{ij}\right)^2&=-3A_{ij}\nabla^2A^{ij}-6A_{ij}R^j_{\;k}A^{ki}-6A_{ij}R^{ikjl}A_{kl}\nn\\
    &=-3A_{ij}\nabla^2A^{ij}+18A_{ij}A^{ij},
\end{align}
where $R_{ikjl}=\gamma_{ij}\gamma_{kl}-\gamma_{il}\gamma_{jk}$ for the unit $S^5$ with metric $\gamma_{ij}$. Therefore
\begin{equation}
    S=\int d^6\xi\sqrt{-g}\;\sum_{i<j}\left(\fft12\dot{A}_{ij}^2+\fft12A_{ij}\left(\nabla^2-6\right)A^{ij}\right).
\end{equation}
In general, transverse $p$-form harmonics $X_{i_1...i_p}$ on $S^d$ satisfy \cite{Iwasaki,Elizalde:1996nb}
\begin{equation}
    \nabla^2 X_{i_1...i_p}=-\left(\ell(\ell+d-1)-p\right)X_{i_1...i_p}.
\end{equation}
There are two transverse two-form harmonics $X^{ij\;(n)}_{(\ell,m_1,m_2,m_3)}$, $n=1,2$ on $S^5$, whose SO$(6)$ representations are labeled by the highest-weight states $[\ell,1,1]$ and $[\ell,1,-1]$.  From (\ref{eq:Dynkin}) we see that the corresponding SU(4) Dynkin labels are $(2,\ell-1,0)$ and $(0,\ell-1,2)$.  These two representations correspond to self-dual and anti-self-dual two forms, respectively, and since the M5-brane world-volume two-form is anti-self-dual, we only take the $(0,\ell-1,2)$ harmonics.

Expanding $A_{ij}$ in anti-self-dual harmonics
\begin{equation}
    A^{ij}(\xi)=\sum_{\ell,m_1,m_2,m_3}A_{(\ell,m_1,m_2,m_3)}(t)X^{ij\;(2)}_{(\ell,m_1,m_2,m_3)}(\xi^k)
\end{equation}
yields
\begin{equation}
    S=\int dt\sum_{\ell,m_a}\left(\fft12\dot A_{(\ell,m_a)}\dot A_{(\ell,-m_a)}-\fft12(\ell+2)^2A_{(\ell,m_a)}A_{(\ell,-m_a)}\right).
\end{equation}
The Hamiltonian is
\begin{equation}
    H=\fft1L\sum_{\ell,m_a}(\ell+2)\left(N_{(\ell,m_a)}+\fft12\right),
\end{equation}
where $N_{(\ell,m_a)}$ are bosonic number operators.  Here we have restored the length dimension $L$ to the Hamiltonian. The charges of the excitations are
\begin{align}
    J&=0,\qquad R_1=0,\nn\\
    R_i&=\sum_{\ell,m_1,m_2,m_3}m_iN_{(\ell,m_1,m_2,m_3)},\qquad i=2,3,4,\nn
\end{align}
The supersymmetric Hamiltonian is then
\begin{equation}
    \mathcal H=\fft12\sum_{\ell,m_1,m_2,m_3}\left((\ell+2-m_1-m_2-m_3)\sum_{n=1,2}N^{(n)}_{(\ell,m_1m_2,m_3)}+(\ell+2)\right),
\end{equation}
and hence the tensor contribution to the single-letter index is
\begin{equation}
    f_{\mathrm{tensor}}=\sum_{\ell,m_j}z^{\ell/2+1}\prod_j(y_j/\sqrt{z})^{m_j}=\sum_{\ell\ge1}z^{\ell/2+1}\chi_{(0,\ell-1,2)}(y_jz^{1/4}/\sqrt{y_2y_3y_4}).
\end{equation}
Performing the sum over $\ell$ gives
\begin{align}
    f_{\mathrm{tensor}}=\fft{z(\lambda\chi_{(0,2)}-\lambda^2\chi_{(0,1)})+z^2(\lambda^{-1}\chi_{(0,1)}-\chi_{(1,1)}-1+\lambda\chi_{(1,0)})}{(1-y_2)(1-y_3)(1-y_4)(1-z/y_2)(1-z/y_3)(1-z/y_4)}\nn\\
    +\fft{z^3(\lambda^{-3}-\lambda^{-2}\chi_{(1,0)}+\lambda^{-1}\chi_{(0,1)}-1)}{(1-y_2)(1-y_3)(1-y_4)(1-z/y_2)(1-z/y_3)(1-z/y_4)},
\label{eq:IM5tensor}
\end{align}
where $\lambda=(y_2y_3y_4)^{1/3}$ and $\chi_{(a,b)}$ are SU(3) characters with fugacities $y_j/\lambda$.   Note that these tensor fluctuations are always gapped, and do not contribute to the index
\begin{equation}
    f_{\mathrm{tensor}}\Big|_{z=0}=0.
\end{equation}
%

%%%%%%%%%%%%%%%%%%%%%%%%%%%%%%%%%%%%%%%%%%%%%
\subsection{Fermion fluctuations}

The fermionic sector of the quadratic action of the M5 brane can be written as \cite{Bandos:1997ui,Beccaria:2023cuo}
\begin{equation}
    S = iT_{M_5} \int d^6\xi \sqrt{-\gamma}\,\bar\theta (1 - \Gamma_{M_5})\Gamma^a \tilde D_a \theta
\end{equation}
where $\theta$ is the $11$ dimensional Majorana spinor,
\begin{align}
    \tilde D_\mu &= \nabla_\mu - \dfrac{1}{288}(\Gamma_\mu^{\ \ \nu\rho\sigma\lambda} + 8 \Gamma^{\rho\sigma\lambda}\delta^\nu_\mu) F_{\nu\rho\sigma\lambda},\qquad F_4 = dA_3,\\
    \Gamma_{M_5} &= \dfrac{1}{6! \sqrt{-\gamma}} \epsilon^{ijklmn}\Gamma_{ijklmn}.
\end{align}
Following the same lines as in the case of the D3-brane, we choose flat world-volume indices to be $(t,\theta_1,\theta_2,\theta_3,\theta_4,\theta_5)\to (0,1,2,3,4,5)$ and $(\zeta_i)\to (6,7,8,9,10)$. Note that in these conventions the AdS$_4$ directions correspond to $(0,6,7,8)$, and the four-form flux is
\begin{equation}
    F_4 = -\dfrac{6}{L} \sqrt{-\gamma}\, \Omega_4,
\end{equation}
where $\Omega_4$ is the volume form on AdS$_4$.

Decomposing the 11-dimensional Dirac matrices according to the conventions in Appendix~\ref{app: Spinors 11 D M5}, we find
\begin{equation}
    \Gamma_{M_5} = \Gamma_{012345} = -\Gamma^{012346} = -\gamma^7
\otimes \mathbf{1}_4=-\sigma^3\otimes\mathbf1_4\otimes\mathbf1_4,
\end{equation}
so the $\kappa$ symmetry projects out spinors based on the parity under $\gamma^7$.  Then we have,
\begin{equation}
    S = iT_{M_5} \int d^6\xi \sqrt{-\gamma}\,\theta^\dagger \left(-\partial_0 + \dfrac{\gamma^0}{L} \fsl\nabla + \dfrac{3}{L} R_1\right)\theta,
\end{equation}
where we have defined, $R_1 = \dfrac{i}{2} \Gamma^{9(10)}= \dfrac{i}{2}\mathbf1_2\otimes\mathbf1_4\otimes\tilde\gamma^{45}$, and where $\theta=\Gamma_{M_5}\theta$.

We now decompose the 11-dimensional spinor $\theta$ according to
\begin{equation}
    \theta=\sum_p\begin{pmatrix}0\\1\end{pmatrix}\otimes\psi_p\otimes\eta_p+\begin{pmatrix}0\\1\end{pmatrix}\otimes\tilde b_5\psi_p^*\otimes\tilde b_5\eta_p^*,
\end{equation}
where $\eta_p$ transforms as the $\mathbf4$ of Sp(4).  The $\psi_p$ can then be further expanded in terms of $S^5$ the spinor harmonics satisfying
\begin{equation}
    \fsl\nabla \chi_{n,\{m_i\}} = \pm i \left(n + \dfrac{5}{2} \right) \chi_{n,\{m_i\}},
\end{equation}
where $ \chi_{n,\{m_i\}}$ transforms under the $[n+1/2,1/2,-1/2]$ and $[n+1/2,1/2,1/2]$ representations of $SO(6)$, or equivalently $(0,n,1)$ and $(1,n,0)$ of SU(4). The $\kappa$ symmetry projects out the later and this decomposition follows the quantum numbers,
\begin{center}
\begin{tabular}{cc|c}
$J$&$R_1$&$H/2$\\
\hline
$+$&$+$&$(n+4)/2$\\
$-$&$-$&$(n+1)/2$\\
$+$&$-$&$(n+1)/2$\\
$-$&$+$&$(n+4)/2$\\
\end{tabular}
\end{center}
where the Hamiltonian is $H L = (n + \fft{5}{2} + 3R_1)$.   In other words, the surviving spinor decomposes as the $\bf4$ of Sp(4) times the $\bar
{\bf4}$ of SU(4).

Summing over the spinor harmonics in the $(0,n,1)$ representation of SU(4) yields the single-letter fermion partition function
\begin{align}
    f_{\mathrm{fermion}}&=\left(\sqrt{p/y_1}+z\left(\sqrt{p y_1} +1/\sqrt{p y_1}\right)+z^2\sqrt{y_1/p}\right)\sum_{\ell,m_j}z^{\ell/2+1/4}\prod_j(y_j/\sqrt{z})^m_j\nn\\
    &=\left(\sqrt{p/y_1}+z\left(\sqrt{p y_1} +1/\sqrt{p y_1}\right)+z^2\sqrt{y_1/p}\right)\sum_{\ell\ge0}z^{\ell/2+1/4}\chi_{(0,\ell,1)}(y_jz^{1/4}/\sqrt{y_2y_3y_4}).
\end{align}
Performing the sum over $\ell$ then gives
\begin{align}
    f_{\mathrm{fermion}}&=\left(\sqrt{p/y_1}+z\left(\sqrt{p y_1} +1/\sqrt{p y_1}\right)+z^2\sqrt{y_1/p}\right)\nn\\
    &\qquad\times\fft{(\lambda^{1/2}\chi_{(0,1)}-\lambda^{3/2})+z(\lambda^{-3/2}-\lambda^{-1/2}\chi_{(1,0)})}{(1-y_2)(1-y_3)(1-y_4)(1-z/y_2)(1-z/y_3)(1-z/y_4)}.
\label{eq:IM5fermion}
\end{align}
The $z=0$ limit gives the fermion contribution to the single-letter index
\begin{equation}
    f_{\mathrm{fermion}}\Big|_{z=0}=\fft{\lambda^2\chi_{(0,1)}-\lambda^3}{(1-y_2)(1-y_3)(1-y_4)},
\end{equation}
where we made use of the relation $p=y_1y_2y_3y_4=y_1\lambda^3$.  Note that the $-\lambda^3$ term in the numerator prevents an overcounting of BPS states in the fermion spectrum.  For example, expanding to $\mathcal O(\lambda^3)$ gives
\begin{align}
    f_{\mathrm{fermion}}\Big|_{z=0}&=(\lambda^2\chi_{(0,1)}-\lambda^3)(1+y_2+y_3+y_4+\cdots)=(\lambda^2\chi_{(0,1)}-\lambda^3)(1+\lambda\chi_{(1,0)}+\cdots)\nn\\
    &=\lambda^2\chi_{(0,1)}+\lambda^3(\chi_{(0,1)}\chi_{(1,0)}-1)+\cdots\nn\\
    &=\lambda^2\chi_{(0,1)}+\lambda^3\chi_{(1,1)}+\cdots.
\end{align}
%

%%%%%%%%%%%%%%%%%%%%%%%%%%%%%%%%%%%%%%%%%%%%%%%
\subsection{Indices from excitations}

The complete single-letter index for a single M5-brane giant graviton can be obtained from the scalar, tensor and fermion contributions, (\ref{eq:IM5scalar}), (\ref{eq:IM5tensor}) and (\ref{eq:IM5fermion}).  When these are added together (with a minus sign for the fermions), the $z$ dependence cancels out as expected, and we are left with the giant graviton index
\begin{equation}
    f_{(1,0,0,0)}(p,y_i)=\fft{p+y_1^{-1}-\lambda^2\chi_{(0,1)}+\lambda^3}{(1-y_2)(1-y_3)(1-y_4)}=\fft{p+y_1^{-1}-y_2y_3y_4(y_2^{-1}+y_3^{-1}+y_4^{-1})+y_2y_3y_4}{(1-y_2)(1-y_3)(1-y_4)}.
\end{equation}
This agrees with (\ref{eq:M5ggindex}).  Note that cancellation of the $z$ dependence between numerator and denominator made use of the identity
\begin{equation}
    (1-z/y_1)(1-z/y_2)(1-z/y_3)=1-z\lambda^{-1}\chi_{(0,1)}+z^2\lambda^{-2}\chi_{(1,0)}+z^3\lambda^{-3},
\end{equation}
which is easily obtained by multiplying out the left-hand-side of the expression.

Table \ref{tab:M5 fluctuations} presents the quantum numbers of all of the M5 fluctuations. The scalars match Table 1 of \cite{Arai:2020uwd}, where the $\mathrm{SO}(3)_J$ vector ${\bf X}$ corresponds to $\zeta_1,\zeta_{2\pm}$ and the singlets $z$, $z^*$ to $\zeta_{3\pm}$. By also including the fermions and tensor, the current Table \ref{tab:M5 fluctuations} goes beyond the analysis of \cite{Arai:2020uwd}, which computed the rest of the index indirectly without the explicit fluctuations.

As in Section \ref{sec:1/2BPS} one may also consider $\frac12$-BPS index in ABJM. Half-BPS operators involve only a single scalar with $H=R_1/2$ and remaining quantum numbers zero \cite{Sheikh-Jabbari:2009vjj}. The expression for the index is exactly the same as \eqref{eq:1/2BPS}, and the giant graviton expansion is also identical \cite{Hayashi:2024aaf}. From Table \ref{tab:M5 fluctuations}, one may verify that the only M5 fluctuation satisfying the $\frac12$-BPS charge relation is $\zeta_{3-,(0,0)}$, which contributes $y_1^{-1}$ to the single-letter index. This is exactly analogous to the D3 brane case, so  
\eqref{eq:halfBPS_PE} and \eqref{eq:1/2BPS_GGE} also apply unchanged.

%%%%%%%%%%%%%%%%%%%%%%%%%%
\begin{table}[t]
    \centering
    \begin{tabular}{c c c c c c c}
        \toprule
        & $H\tilde L$ & $J$ & $R_1$ & $[R_2,R_3,R_4]$ & $\mathcal{H}$\\
        \midrule
        $\zeta_{1,(\ell,m_j)}$& $(\ell+2)/2$ & $0$ & $0$ & $[\ell,0,0]$ & $(\ell+2-m_1-m_2-m_3)/2$\\
        $\zeta_{2+,(\ell,m_j)}$& $(\ell+2)/2$ & $1$ & $0$ & $[\ell,0,0]$& $(\ell-m_1-m_2-m_3)/2$ \\
        $\zeta_{3+,(\ell,m_j)}$& $(\ell+5)/2$ & $0$ & $1$ & $[\ell,0,0]$ & $(\ell+3-m_1-m_2-m_3)/2$ \\
        $\zeta_{2-,(\ell,m_j)}$& $(\ell+2)/2$ & $-1$ & $0$ & $[\ell,0,0]$ & $(\ell+4-m_1-m_2-m_3)/2$\\
        $\zeta_{3-,(\ell,m_j)}$& $(\ell-1)/2$ & $0$ & $-1$ & $[\ell,0,0]$ & $(\ell - m_1 - m_2 - m_3)/2$ \\
        \midrule
        $A_{\ell,m_j}$ & $(\ell + 2)/2$ & $0$ &$0$ & $[\ell,1,-1]$ & $(\ell + 2 - m_1 - m_2 - m_3)/2$\\
        \midrule
        $\psi_{++,(\ell,m_j)}$ & $(\ell+4)/2$ & $1/2$ & $1/2$ & $[\ell+1/2,1/2,-1/2]$ & $(2\ell+5-2m_1-2m_2-2m_3)/4$\\
        $\psi_{--,(\ell,m_j)}$ & $(\ell+1)/2$ & $-1/2$ & $-1/2$ & $[\ell+1/2,1/2,-1/2]$ & $(2\ell + 5 -2m_1-2m_2-2m_3)/4$ \\
        $\psi_{+-,(\ell,m_j)}$ & $(\ell+1)/2$ & $1/2$ & $-1/2$ & $[\ell+1/2,1/2,-1/2]$ & $(2\ell+1-2m_1-2m_2-2m_3)/4$ \\
        $\psi_{-+,(\ell,m_j)}$ & $(\ell+4)/2$ & $-1/2$ & $1/2$ & $[\ell+1/2,1/2,-1/2]$ & $(2\ell+9-2m_1-2m_2-2m_3)/4$ \\
        \bottomrule
    \end{tabular}
    \caption{Quantum numbers of the M5 brane fluctuations corresponding to each of the letters. We have subtracted the ground state and classical contributions to $H$. Here $[R_2,R_3,R_4]$ labels the $SO(6)$ representations with the charges of their highest weight states and $m_1,m_2,m_3$ are the charges of a general state in that representation. The $\pm$ subscripts on the spinors corresponds to their $J, R_1$ charges.}
    \label{tab:M5 fluctuations}
\end{table}
%%%%%%%%%%%%%%%%%%%%%%%%%%%%%

%%%%%%%%%%%%%%%%%%%%%%%%%%%%%%%%%%%%%%%%%%%%%%%
\section{M2 giant graviton in \texorpdfstring{AdS$_7\times S^4$}{AdS7xS4}}\label{Sec:M2}

Finally, we consider the six-dimensional $\mathcal N=(2,0)$ index which corresponds to the M5-brane world-volume dual to AdS$_7\times S^4$.  The six-dimensional $\mathcal{N}=(2,0)$ superconformal algebra $OSp(8^*|4)$ has bosonic subalgebra $SO(2,6)\times SO(5)$ with six Cartan generators
\begin{equation}
    H,\quad J_1,\quad J_2,\quad J_3,\quad R_1,\quad R_2
\end{equation}
corresponding to the Hamiltonian, three angular momenta and two R-charges, respectively. The superconformal index may be defined as \cite{Arai:2020uwd}
\begin{equation}
    \mathcal{I}(p_i;y_j)=\operatorname{Tr}\left[(-1)^F e^{-\beta\mathcal H} p_1^{J_1}p_2^{J_2}p_3^{J_3} y_1^{R_1}y_2^{R_2}\right],\qquad p_1p_2p_3=y_1y_2,
\end{equation}
where
\begin{equation}
    \mathcal H=H-J_1-J_2-J_3-2(R_1+R_2),
\label{eq:M5sciH}
\end{equation}
and its giant graviton expansion takes the form
\begin{equation}
    \fft{\mathcal I_N(p_i;y_j)}{\mathcal I_\infty(p_i;y_j)}=\sum_{m_1,m_2\ge0}y_1^{m_1N}y_2^{m_2N}\hat I^{\mathrm{GG}}_{(m_1,m_2)}(p_i;y_j).
\end{equation}
Here $m_1$ and $m_2$ correspond to the number of M2-branes wrapped on $S^2\subset S^4$ and moving along the rotation planes corresponding to $R_1$ and $R_2$, respectively.

The giant graviton index $\hat I^{\mathrm{GG}}_{(m_1,m_2)}$ is identified as a multiple M2-brane index $\mathcal I_{M2}$.  For a single giant graviton, we have $\hat I^{\mathrm{GG}}_{(1,0)}(p_1,p_2,p_3;y_1,y_2)=\mathcal I_{M2}(p,y_1,y_2,y_3,y_4)$ with fugacities mapped according to
\begin{equation}
    p\to y_2,\quad y_1\to y_1^{-1},\quad y_2\to p_1,\quad y_3\to p_2,\quad y_4\to p_3.
\end{equation}
In particular, $\hat I^{\mathrm{GG}}_{(1,0)}$ is the plethystic exponential of the single-letter index
\begin{equation}
    f_{(1,0)}(p_i;y_j)=\fft{y_1^{-1}+p_1+p_2+p_3-y_2(y_1+p_1^{-1}+p_2^{-1}+p_3^{-1})}{1-y_2}.
\label{eq:M2ggf}
\end{equation}
We now demonstrate how this can be obtained from the quantized M2-brane fluctuations.

%%%%%%%%%%%%%%%%%%%%%%%%%
\subsection{Scalar fluctuations}

The world-volume theory of a stack of $N$ M2-branes on a $\mathbb Z_k$ orbifold is  generically an $\mathcal N=6$ Chern-Simons-matter theory known as ABJM theory.  However, a single M2-brane can be described by a free $\mathcal N=8$ theory with eight scalars and eight fermions.

For the scalar fluctuations, the supersymmetric Hamiltonian, (\ref{eq:M5sciH}),  is given by
\begin{align}\label{eq:M2_susyH}
    \mathcal H&=\left(H\tilde{L}-2N\right)-\sum_{i=1}^3 J_i-2\sum_{i=1}^2 R_i\nn\\
    &=2\sum_{\ell,m}\biggl(\left(\ell- m\right)\left(N_{1+,(\ell,m)}+N_{2+,(\ell,m)}+N_{3+,(\ell,m)}+N_{4-,(\ell,m)}\right)\nn\\
    &\kern4em+\left(\ell-m+1\right)\left(N_{1-,(\ell,m)}+N_{2-,(\ell,m)}+N_{3-,(\ell,m)}+N_{4+,(\ell,m)}\right)+4\left(\ell+\fft12\right)\biggr).
\end{align}
Here the modes on $S^2$ are expanded in ordinary spherical harmonics $Y_{\ell,m}$, and the sum over $m$ yields a SU(2) character $\chi_\ell$.  As a result, the single-letter scalar partition function is
\begin{align}
    f_\text{scalar}&=\left(p_1+p_2+p_3+y_1^{-1}+z(p_1^{-1}+p_2^{-1}+p_3^{-1}+y_1)\right)\sum_{\ell\geq0}z^\ell\chi_\ell(y_2/z)\nn\\
    &=\fft{\left(p_1+p_2+p_3+y_1^{-1}+z(p_1^{-1}+p_2^{-1}+p_3^{-1}+y_1)\right)(1+z)}{(1-y_2)(1-y_2^{-1}z^2)},
\label{eq:M2scalar}
\end{align}
and the contribution to the index is simply
\begin{equation}
    f_\text{scalar}\Big|_{z=0}=\fft{p_1+p_2+p_3+y_1^{-1}}{1-y_2}.
\end{equation}
This last result may also be derived by noting that only excitations from the second line of \eqref{eq:M2_susyH} with $m=\ell$ are BPS. The excitations $N_{1+,(\ell,\ell)}$, $N_{2+,(\ell,\ell)}$, $N_{3+,(\ell,\ell)}$ and $N_{4-,(\ell,\ell)}$ contribute $p_1$, $p_2$, $p_3$ and $y_1^{-1}$ to the index respectively, with the denominator coming from a geometric sum over all $\ell$.

%%%%%%%%%%%%%%%%%%%%%%%%%%%%%%%%%%%%%%%%%%
\subsection{Fermion fluctuations}
From \cite{Marolf:2004jb} the action for the fermionic sector takes the form,
\begin{equation}
    S = \dfrac{ i T_{M_2}}{2} \int d^3\xi \sqrt{-\gamma}\, \bar \theta (1 - \Gamma_{M_2}) \Gamma^a \tilde{D}_a \theta
\end{equation}
where $\theta$ is an $11$ dimensional Majorana spinor,
\begin{align}
    \tilde D_\mu &= \nabla_\mu- \dfrac{1}{288} \Gamma^\mu(\Gamma_\mu^{\;\;\nu\rho\sigma\lambda} - 8 \delta_\mu^\nu \Gamma^{\rho\sigma\lambda})F_{\nu\rho\sigma\lambda},\qquad F_4 = dA_3\\
    \Gamma_{M_2} &=\dfrac{1}{3!\sqrt{-\gamma}}\epsilon^{ijk}\Gamma_{ijk}.
\end{align}

We choose the $M_2$ brane world-volume tangent space indices to be $(t,\theta_1\theta_2)\to (0,1,2)$ and $(\zeta_i)\to (3,4,5,6,7,8,9,10)$ so that
\begin{equation}
    F_4 = - \dfrac{3}{L} \sqrt{-\gamma}\, \Omega_4,
\end{equation}
where $\Omega_4$ is the volume form on $S^4$. Following the conventions in Appendix~\ref{app: Spinors 11 D M2}, we take the Dirac decomposition
\begin{equation}
    \Gamma^0 = i\sigma_2 \otimes \tilde\gamma^9, \quad \Gamma^1 = \sigma_1\otimes \tilde\gamma^9, \quad \Gamma^2 = \sigma_3 \otimes\tilde\gamma^9,\quad \Gamma^i = \mathbf 1_2 \otimes \tilde\gamma^{i-2}.
\end{equation}
Then we can write
\begin{equation}
    - \dfrac{1}{288}\Gamma^\mu(\Gamma_\mu^{\;\;\nu\rho\sigma\lambda} - 8 \delta_\mu^\nu \Gamma^{\rho\sigma\lambda})F_{\nu\rho\sigma\lambda} = \dfrac{3}{4L} i\sigma_2 \otimes \tilde\gamma^{78},
\end{equation}
and
\begin{equation}
    \Gamma_{M_2} = \Gamma_{012} = -\Gamma^{012} = - \mathbf 1_2 \otimes \tilde\gamma^9.
\end{equation}
Since $\kappa$ symmetry demands $\bar\theta \Gamma_{M_2} = - \bar\theta$ (or $\Gamma_{M_2}\theta = -\theta$), in our basis this selects spinors that have $+$ parity under $\tilde\gamma^9$. In addition to this, the Majorana condition demands that $B_{11}\theta = \theta^*$.
Thus we decompose the spinor $\theta$ as
\begin{equation}
    \theta = \sum_p \psi_p \otimes\eta^{(+)} + \psi_p^* \otimes \tilde b_8 \eta^{(+)}.
\end{equation}
Then we have,
\begin{equation}
    S = \dfrac{iT}{L} \sum_p \int dt\, d^2\xi L^2 \sqrt{\gamma} \, \psi^\dagger\left(-\partial_0 + \fsl\nabla + \dfrac{3i}{2} R_1\right) \psi + c.c.,
\end{equation}
where $R_1 = \fft{i}{2}\gamma^{78}$ and $\fsl\nabla = \sigma^3 \nabla_1 - \sigma^1\nabla_2$. Expanding the spinor in spinor harmonics on $S^2$,
\begin{equation}
    \psi_p = \psi_{p,(n,m)} \chi_{(n,m)},
\end{equation}
one has,
\begin{equation}
    \fsl\nabla \chi_{(n,m)} = i(n+1) \chi_{(n,m)},
\end{equation}
where $\chi_{(n,m)}$ are in the $n + \fft{1}{2}$ representation of $su(2)$. Then the Hamiltonian takes the form,
\begin{equation}
    H = \dfrac{2}{\tilde L} \sum_{p,n,m} \left(n + 1 + \dfrac{3}{2}R_1\right) \psi_{p,(n,m)}^\dagger \psi_{p,(n,m)}.
\end{equation}
To summarise, the $11$ dimensional Majorana spinors can be decomposed into a set of three-dimensional spinors $\psi_p$ attached to the spinors $\eta^{(\pm)}$ transforming as the $\mathbf 8_+$ and $\mathbf 8_-$ of $SO(8)$. Since $\mathbf 8_-$ have $-$ parity, it will be projected out by $\kappa$ symmetry. So only the $\mathbf 8_+$ will contribute. These will be further decomposed to the $n + \fft{1}{2}$ representations of $SU(2)$ according to
\begin{center}
\begin{tabular}{cccccc|c|c}
{}&$J_1$&$J_2$& $J_3$&$R_1$&$R_2$&$2H$&$\Delta$\\
\hline
{}&$+$&$+$&$+$&$+$&$m+1/2$&$2n+7/2$&$2n-2m$\\
$3\times$&$+$&$+$&$-$&$-$&$m+1/2$&$2n+1/2$&$2n-2m$\\
$3\times$&$+$&$-$&$-$&$+$&$m+1/2$&$2n+7/2$&$2n-2m+2$\\
{}&$-$&$-$&$-$&$-$&$m+1/2$&$2n+1/2$&$2n-2m+2$\\
\hline
\\\end{tabular}
\end{center}
where the notation is that the middle entries have $J_1$, $J_2$ and $J_3$ cyclically permuted.  From the table one can write down the single-letter partition function as,
\begin{align}
    f_{\mathrm{fermion}} &= \biggl((p_1p_2p_3y_1)^{1/2} + \left(\dfrac{p_1p_2}{p_3y_1}\right)^{1/2}+\left(\dfrac{p_2p_3}{p_1y_1}\right)^{1/2}+\left(\dfrac{p_3p_1}{p_2y_1}\right)^{1/2}\nn\\
    &\kern4em+z\biggl((p_1p_2p_3y_1)^{-1/2}+\left(\fft{p_3y_1}{p_1p_2}\right)^{1/2}+\left(\fft{p_1y_1}{p_2p_3}\right)^{1/2}+\left(\fft{p_2y_1}{p_3p_1}\right)^{1/2}\biggr)\biggr)\nn\\
    &\qquad\times\sum_{\ell=0}^\infty z^{\ell+1/2}\chi_{\ell+1/2}(y_2/z)\\
    &= \dfrac{(y_2(y_1+p^{-1}_1+p^{-1}_2+p^{-1}_3)+z(y_1^{-1}+p_1+p_2+p_3))(1+y_2^{-1}z)}{(1 - y_2)(1-y_2^{-1}z^2)},
\label{eq:M2fermion}
\end{align}
where in the second line we have used $p_1p_2p_3 = y_1y_2$.  The corresponding contribution to the index is
\begin{equation}
    f_{\mathrm{fermion}}\Big|_{z=0}=\fft{y_2(y_1+p^{-1}_1+p^{-1}_2+p^{-1}_3)}{1-y_2}.
\end{equation}

%%%%%%%%%%%%%%%%%%%%%%%%%%%%%%%%%%%%%%%%%%%%%%%
\subsection{Indices from excitations}

Combining the scalars (\ref{eq:M2scalar}) and fermions (\ref{eq:M2fermion}) gives the single-letter giant graviton index
\begin{equation}
    f_{(1,0)}(p_i;y_j)=\fft{y_1^{-1}+p_1+p_2+p_3-y_2(y_1+p_1^{-1}+p_2^{-1}+p_3^{-1})}{1-y_2},
\end{equation}
in agreement with (\ref{eq:M2ggf}). All modes, both BPS and non-BPS, are presented in Table \ref{tab:M2 fluctuations}. Note that the scalars $\zeta_{1\pm}$, $\zeta_{2\pm}$, and $\zeta_{3\pm}$ transform under the vector representation of the $\rm{SO}(6)_J$ angular momentum group. Also, fermions of the form $\psi_{\pm\pm\pm+}$ and $\psi_{\pm\pm\pm-}$ organize into the $\mathbf{4}$ and $\mathbf{\bar 4}$ respectively. Table 2 matches \cite{Arai:2020uwd}, but now with the fermions included.

As for the D3 and M5 cases, one may derive the giant graviton expansion for the $\frac12$-BPS index. It again takes the same form as \eqref{eq:1/2BPS_GGE}, as shown in \cite{Hayashi:2024aaf}, and counts only fluctuations $\zeta_{4-,(0,0)}$.

\begin{table}[ht]
    \centering
    \begin{tabular}{c c c c c c c c}
        \toprule
        {}&$H\tilde L$ & $J_1$ & $J_2$ & $J_3$ & $R_1$ & $[R_2]$ & $\mathcal{H}$\\
        \midrule
        $\zeta_{1+,(\ell,m_1)} $ & $2\ell + 1$ & $1$ & $0$ & $0$ & $0$ & $[\ell]$ & $2\ell - 2m_1$ \\
        $\zeta_{2+,(\ell,m_1)} $ & $2\ell + 1$ & $0$ & $1$ & $0$ & $0$ & $[\ell]$ & $2\ell - 2m_1$ \\
        $\zeta_{3+,(\ell,m_1)} $ & $2\ell + 1$ & $0$ & $0$ & $1$ & $0$ & $[\ell]$ & $2\ell - 2m_1$ \\
        $\zeta_{4+,(\ell,m_1)} $ & $2\ell + 4$ & $0$ & $0$ & $0$ & $1$ & $[\ell]$ & $2\ell + 2 - 2m_1$ \\
        $\zeta_{1-,(\ell,m_1)} $ & $2\ell + 1$ & $-1$ & $0$ & $0$ & $0$ & $[\ell]$ & $2\ell + 2 - 2m_1$ \\
        $\zeta_{2-,(\ell,m_1)} $ & $2\ell + 1$ & $0$ & $-1$ & $0$ & $0$ & $[\ell]$ & $2\ell + 2 - 2m_1$ \\
        $\zeta_{3-,(\ell,m_1)} $ & $2\ell + 1$ & $0$ & $0$ & $-1$ & $0$ & $[\ell]$ & $2\ell + 2 - 2m_1$ \\
        $\zeta_{4-,(\ell,m_1)} $ & $2\ell - 2$ & $0$ & $0$ & $0$ & $-1$ & $[\ell]$ & $2\ell - 2m_1$ \\
        \midrule
        $\psi_{++++,(\ell,m_1)}$ & $2\ell + 7/2$ & $1/2$ & $1/2$ & $1/2$ & $1/2$ & $[\ell + 1/2] $ & $2\ell + 1 - 2m_1$\\
        $\psi_{+--+,(\ell,m_1)}$ & $2\ell + 7/2$ & $1/2$ & $-1/2$ & $-1/2$ & $1/2$ & $[\ell + 1/2] $ & $2\ell + 3 - 2m_1$\\
        $\psi_{-+-+,(\ell,m_1)}$ & $2\ell + 7/2$ & $-1/2$ & $1/2$ & $-1/2$ & $1/2$ & $[\ell + 1/2] $ & $2\ell + 3 - 2m_1$\\
        $\psi_{--++,(\ell,m_1)}$ & $2\ell + 7/2$ & $-1/2$ & $-1/2$ & $1/2$ & $1/2$ & $[\ell + 1/2] $ & $2\ell + 3 - 2m_1$ \\
        $\psi_{-++-,(\ell,m_1)}$ & $2\ell + 1/2$ & $-1/2$ & $1/2$ & $1/2$ & $-1/2$ & $[\ell + 1/2] $ & $2\ell + 1 - 2m_1$\\
        $\psi_{+-+-,(\ell,m_1)}$ & $2\ell + 1/2$ & $1/2$ & $-1/2$ & $1/2$ & $-1/2$ & $[\ell + 1/2] $ & $2\ell + 1 - 2m_1$\\
        $\psi_{++--,(\ell,m_1)}$ & $2\ell + 1/2$ & $1/2$ & $1/2$ & $-1/2$ & $-1/2$ & $[\ell + 1/2] $ & $2\ell + 1 - 2m_1$\\
        $\psi_{----,(\ell,m_1)}$ & $2\ell + 1/2$ & $-1/2$ & $-1/2$ & $-1/2$ & $-1/2$ & $[\ell + 1/2] $ & $2\ell + 3 - 2m_1$\\
        \bottomrule
    \end{tabular}
    \caption{Quantum numbers of the M2 brane fluctuations corresponding to each letter. We have subtracted the ground state and classical contributions to $H$. Here, $[R_2]$ labels the $SO(3)$ representations by the charge of their highest weight states, and $m_1$ is the charge of a general state in that representation. The $\pm$ subscripts on the spinors corresponds to their $J_1, J_2,J_3, R_1$ charges.}
    \label{tab:M2 fluctuations}
\end{table}

%%%%%%%%%%%%%%%%%%%%%%%%%%%%%%%%%%%%%%%%%%%%%%%%%%%
\section{The full \texorpdfstring{$\frac{1}{2}$-BPS}{1/2-BPS} index from  non-Abelian DBI}\label{Sec:nonAb}

In this section, we generalize our discussion of scalar fluctuations to a stack of multiple giants. Despite the absence of a full non-Abelian generalization of the DBI action, one may still use a matrix description to obtain quadratic fluctuations for the case of coincident D3 branes in AdS$_5\times S^5$. This will be sufficient to derive the giant graviton expansion of the $\frac12$-BPS index to all orders. At quadratic level, non-Abelianization of the D3 brane action amounts to promoting fields to adjoint matrices. To justify this result, at least for the bosonic sector, one may consider the DBI generalization described in \cite{Myers:2003bw} (see also \cite{Tseytlin:1997csa,Tseytlin:1999dj} for background). The argument is given in Appendix \ref{app:nA DBI}. Once we restrict the action presented in Equations (12) and (13) of \cite{Myers:2003bw} to quadratic order in the scalars, one essentially obtains the single-brane fluctuation action but with the coordinates as matrices.

We shall specialize to the case of D3 branes, since a simple matrix generalization is not expected for the M2/M5 brane action. However, the $\frac12$-BPS indices in all three theories are exactly analogous, so there may be some validity in extending the argument below to the M2 and M5 cases.

Recall the giant graviton expansion of the $\frac{1}{2}$-BPS index in \eqref{eq:1/2BPS_GGE}. There we considered only fluctuations of a single D3 brane, which resulted in just the first nontrivial term in the expansion. Promoting the scalar fluctuations to $m\times m$ matrices will produce the $y_1^{mN}$ term in the $\frac{1}{2}$-BPS index. To derive the non-Abelian action, we first restrict \eqref{eq:S} to only the $\frac{1}{2}$-BPS modes. Such modes must satisfy $H=R_1$ with all remaining charges and angular momenta zero. From Table \ref{tab: D3 fluctuations}, only the $s$-wave part of one scalar mode 
\begin{equation}
    \zeta_{3-,(0,0,0)}=\fft1{\sqrt{2}}(\zeta_{5,(0,0,0)}-i\zeta_{6,(0,0,0)})
\end{equation} contributes. Restricting the Hamiltonian \eqref{eq:scalarH} and $R_1$ charge \eqref{eq:scalarR1} to only $x\equiv \zeta_{5,(0,0,0)}$ and $y\equiv \zeta_{6,(0,0,0)}$ gives
\begin{align}
    H&=N+\fft12\left(p_x^2+p_y^2\right)+\fft12(x^2+y^2)+2(xp_y-yp_x)\nn\\
    R_1&=xp_y-yp_x.
\end{align}
As before, we define the supersymmetric Hamiltonian $\mathcal{H}\equiv H-N-R_1$. Promoting these quantities to operators and subtracting the zero-point energy yields
\begin{align}\label{eq:zHam}
    \mathcal{H}&=-\frac{\partial^2}{\partial z\partial\bar{z}}+z\bar{z}+z\pd[]{}{z}+\bar{z}\pd[]{}{\bar{z}}-1\nn\\
    R_1&=z\pd[]{}{z}+\bar{z}\pd[]{}{\bar{z}},
\end{align}
where we have introduced complex coordinates,
\begin{align}
    z\equiv\frac{x+iy}{\sqrt{2}},\qquad \pd[]{}{z}\equiv\frac1{\sqrt{2}}\left(\pd[]{}{x}-i\pd[]{}{y}\right).
\end{align}
Before studying the non-Abelian generalization of this system, it is useful to re-derive the single-brane index \eqref{eq:halfBPS_PE} using the wavefunction description of fluctuations. Wavefunctions satisfying $\mathcal{H}\Psi(z,\bar{z})=0$ can be expressed using the basis functions
\begin{equation}\label{eq:zWave}
    \Psi_n(z,\bar{z})\propto\bar{z}^ne^{-z\bar{z}}, \qquad n\in\mathbb{N},
\end{equation}
which each carry $-n$ units of $R_1$ charge:
\begin{equation}
    R_1\Psi_n(z,\bar{z})=-n\Psi_n(z,\bar{z}).
\end{equation}
Evaluating the index for $\frac{1}{2}$-BPS fluctuations yields
\begin{equation}
    \hat I^{\mathrm{GG}}_\text{1-giant}(y_1) = \underset{\mathcal{H}=0}{\Tr} y_1^{R_1}=\sum_{n=0}^\infty y_1^{-n}=\frac1{1-y_1^{-1}},
\end{equation}
which agrees with \eqref{eq:halfBPS_PE}.

To extend these results to a stack of multiple giants, we  promote the coordinate $z$ in \eqref{eq:zHam} to an $m\times m$ matrix $Z_{ij}$.
\begin{align}\label{eq:non_ab_H}
    \mathcal{H}&=\sum_{i,j=1}^m\left(-\frac{\partial^2}{\partial Z_{ij}\partial \overline{Z_{ij}}}+Z_{ij}\overline{Z_{ij}}+Z_{ij}\pd[]{}{Z_{ij}}+\overline{Z_{ij}}\pd[]{}{\overline{Z_{ij}}}-1\right)\nn\\
    R_1&=\sum_{i,j=1}^m\left(Z_{ij}\pd[]{}{Z_{ij}}+\overline{Z_{ij}}\pd[]{}{\overline{Z_{ij}}}\right).
\end{align}
The ground state wavefunction is
\begin{equation}
    \Psi_0(Z,\overline{Z})=\exp\left(-\sum_{i,j=1}^m Z_{ij}\overline{Z_{ij}}\right)=\exp\left(-\Tr ZZ^\dagger\right).
\end{equation}
Similar to \eqref{eq:zWave}, excited wavefunctions may be constructed by multiplying the ground state by any $\overline{Z_{ij}}$. However, $U(m)$ invariance of the system requires all indices to be contracted. The resulting wavefunctions take the form of multitraces
\begin{equation}\label{eq:multitraces}
    \Psi_{r_1,r_2,\dots r_m}\left(Z,Z^\dagger\right)\quad=\quad\Tr\left(Z^{\dagger }\right)^{r_1}\Tr\left(Z^{\dagger 2}\right)^{r_2}\cdots\Tr\left(Z^{\dagger m}\right)^{r_m}\exp\left(-\Tr ZZ^\dagger\right),
\end{equation}
and are parameterized by integers $r_1,r_2,\dots r_m$. Due to trace relations, any $\Tr{(Z^{\dagger k})}$ with $k>m$ can be expressed as a linear combination of traces of lower powers, and thus does not need to be considered. The $R_1$ charge is determined by the total number of $Z^\dagger$'s:
\begin{equation}
    R_1\Psi_{r_1,r_2,\ldots r_m}\left(Z,Z^\dagger\right)=-(r_1+2r_2+3r_3\cdots+mr_m)\Psi_{r_1,r_2,\ldots r_m}\left(Z,Z^\dagger\right).
\end{equation}
The giant graviton index is given by the generating function for these multitraces,
\begin{equation}\label{eq:1/2BPS_I_hat}
    \hat I^{\mathrm{GG}}_m(y_1)=\prod_{n=1}^m\frac1{1-y_1^{-n}},
\end{equation}
which exactly matches \eqref{eq:1/2BPS_GGE}. Each term in the product is a geometric series accounting for any power of $\Tr(Z^{\dagger n})$, and the full expression combines traces of different $n$. Note that this derivation closely parallels the direct field theory computation of the $\frac12$-BPS index, which counts multitraces of a single complex scalar $Z$, in $\mathcal{N}=4$ SYM. The only difference is that the wavefunctions \eqref{eq:multitraces} contain traces of $Z^\dagger$ instead of $Z$, which flips the sign of their R charge.

As in \cite{Takayama:2005yq,Eleftheriou:2023jxr}, one may also derive the same result in terms of matrix eigenvalues. One gauge-fixes the overall $U(m)$ symmetry by rotating $Z$ into upper triangular form. The degrees of freedom reduce to the eigenvalues $z_i$, and the wavefunctions may be written in terms of symmetric polynomials in $z_i$. Counting polynomials in an appropriate basis, e.g. the Schur basis, also yields \eqref{eq:1/2BPS_I_hat}.

The enumeration of the wavefunctions \eqref{eq:multitraces} may be interpreted a concrete realization of the integral expression for $\hat I^{\mathrm{GG}}_m(y_1)$ appearing in \cite{Lee:2022vig}:

\begin{equation}\label{eq:Im_CI}
    \hat I^{\mathrm{GG}}_m(y_1)=\frac{1}{m!} \oint \prod_a \frac{d \sigma_a}{2 \pi i \sigma_a} \prod_{a \neq b}\left(1-\sigma_a / \sigma_b\right) \prod_{a, b} \frac{1}{\left(1-y_1^{-1} \sigma_a / \sigma_b\right)}.
\end{equation}
Here the single-letter index for one giant, $y_1^{-1}$, has been multiplied by the adjoint character $\sigma_a / \sigma_b$. This effectively promotes the scalar fluctuation of the giant to a matrix. The final product in \eqref{eq:Im_CI} originates from the plethistic exponential of $y_1^{-1}\sigma_a / \sigma_b$, and the contour integral with the Haar measure projects out the $\textrm{U}(m)$ singlets. This precisely encodes the form of the wavefunctions in \eqref{eq:multitraces}, which consist of products of $Z^\dagger$ with all indices contracted via multitraces.

\section{Conclusions}\label{Sec:Conclusions}
In this manuscript we have systematically derived the spectrum of fluctuations of giant gravitons as probes in supergravity backgrounds. We have quantized those fluctuations and demonstrated that they precisely generate the single-letter index of the giant graviton expansions for the respective dual field theories. Namely, the D3 brane giant graviton generates the single-letter indices for ${\cal N}=4$ SYM, the M5 brane the indices for ABJM theory and the M2 brane generates indices for the 6d $(2,0)$ theory. We were also able to derive the full finite $N$ giant graviton expansion for the $\frac{1}{2}$-BPS index of ${\cal N}=4$ SYM by extending certain Abelian fluctuations to matrix-valued fields.

There have been some previous works exploring the role of giant graviton fluctuations in the expansions of various indices. For example, the works \cite{Beccaria:2023cuo,Beccaria:2024vfx,Gautason:2024nru} were able to reproduce the one-graviton contribution, $q^N$, to the index. We have focused on quantizing all the fluctuations and understanding how they combine to yield the single-letter indices. 

The broader framework of  open-closed-open triality discussed recently in \cite{Gopakumar:2024jfq} provides a conceptual setup for the discussion in this manuscript. The three descriptions are complete at the level of one giant graviton when the discussion is Abelian. The crucial open technical question is whether the giant graviton point of view can be made precise for states  beyond the $\frac12$-BPS sector. In particular, it would be substantial progress to understand the $\frac{1}{4}$- and $\frac{1}{8}$-BPS sectors directly in the giant graviton open string framework. 

It would be interesting to explore other classes of theories, including AdS$_5\times SE_5$ with known ${\cal N}=1$ toric quiver gauge theories. For these theories the evaluation of the superconformal index is known \cite{Cabo-Bizet:2019osg, Kim:2019yrz,GonzalezLezcano:2019nca, Lanir:2019abx, GonzalezLezcano:2020yeb} and an understanding of the giant gravitons might help in clarifying properties of gravity that are universal in asymptotically AdS spacetimes. Similarly,  it would be interesting to extend our derivations of the full spectrum of fluctuations and their quantizations to non-conformal Dp branes recently considered in \cite{Batra:2025ivy}.

The most ambitious goal and formidable obstacle remains the description of multiple giant gravitons, which requires a general non-Abelian  action as well as a description of modes at brane intersections. This seems to be the clear obstruction to obtaining finite-$N$ results. In a sense, {\it we  have the atoms of supersymmetric configurations but, unfortunately,  cannot form molecules yet.} Along these lines, it would be quite interesting to explore simpler situations where non-Abelianization can be achieved by other means such as in the case of the $\frac{1}{2}$-BPS index discussed in \cite{Lee:2023iil,Eleftheriou:2023jxr} and in this paper. We hope to return to some of these interesting questions elsewhere.

%%%%%%%%%%%%%%%%%%%%%%%%%%%%%%%%%
\acknowledgments
We are thankful to Diego Correa, Ignacio Cruz, Alberto Faraggi, Alfredo Gonz\'alez Lezcano, Henry Lin, Sameer Murthy, Augniva Ray and Diego Trancanelli for clarifying discussions or ongoing explorations in related topics. This work is partially supported by the U.S. Department of Energy under grant DE-SC0007859. E.D. and S.J. acknowlege support from the Leinweber Institute for Theoretical Physics in the form of Summer Fellowships.

%%%%%%%%%%%%%%%%%%%%%%%%%%%%%%%%
\appendix

\section{Spinor conventions}
In this section we discus the conventions used for the decomposition of the higher dimensional spinors to corresponding giant graviton world-volume spinors. For the D3 brane giant graviton of $AdS_5 \times S^5$, we decompose the $10$ dimensional Majorana-Weyl Spinor into $SO(1,3)\times SO(6)$ spinors. For the M2 giants of $AdS_7 \times S_4$, we decompose the $11$ dimensional Majorana spinor into $SO(1,2) \times SO(8)$ spinors and for the M5 giants of $AdS_4 \times S_7$ we decompose this into $SO(1,5) \times SO(5)$ spinors.

Our spinor conventions mostly follow that of \cite{Freedman:2012zz}, but with general spacetime signatures treated as in \cite{Kugo:1982bn}.  Since we consider both Euclidean and Lorentzian spinors, it is useful to review some basic properties of the Dirac matrics in general dimensions with arbitrary signature, $D=t+s$.  We take the Clifford algebra
\begin{equation}
    \{\Gamma^\mu,\Gamma^\nu\}=2\eta^{\mu\nu}.
\end{equation}
where $\eta^{\mu\nu}$ has $t$ negative eigenvalues and $s$ positive ones.  In particular, $\Gamma^0$ through $\Gamma^{t-1}$ are anti-Hermitian while $\Gamma^t$ through $\Gamma^{t+s-1}$ are Hermitian.  We define the three matrices $A$, $B$ and $C$ according to
\begin{align}
    \mbox{Hermitian conjugation:}&&\Gamma_\mu^\dagger&=(-1)^tA\Gamma_\mu A^{-1},\qquad&& A=A^\dagger=A^{-1},\nn\\
    \mbox{Charge conjugation:}&&\Gamma_\mu^T&=\eta(-1)^tC\Gamma_\mu C^{-1},\qquad&& CC^\dagger=1,\nn\\
    \mbox{Majorana conjugation:}&&\Gamma_\mu^*&=\eta B\Gamma_\mu B^{-1},\qquad&& B=CA,\quad BB^\dagger=1.
\end{align}
Here $\eta=\pm1$ is a choice of sign with both choices possible in even dimensions, but with only one possibility in odd dimensions.  A second choice of sign, $\epsilon$, exists for the transpose of the charge conjugation matrix
\begin{equation}
    C^T=\epsilon\eta^t(-1)^{\fft{t(t+1)}2}C.
\end{equation}
If we consider a representation where each Dirac matrix is either real or imaginary and either symmetric or antisymmetric, it is easy to convince ourselves that, up to some phases, $A$ is the product of the anti-Hermitian Dirac matrices
\begin{equation}
    A=i^{\fft{t(t+1)}2}\Gamma_0\cdots\Gamma_{t-1},
\end{equation}
while $B$ is the product of all imaginary or of all real Dirac matrices and $C$ is the product of all symmetric or of all antisymmetric Dirac matrices.

The Majorana and Dirac spinor conjugates are
\begin{equation}
    \psi^c=\psi^TC,\qquad\bar\psi=\psi^\dagger A,
\end{equation}
and the Majorana condition, $\psi^c=\bar\psi$, becomes simply
\begin{equation}
    \psi^*=B\psi.
\end{equation}
This can only be satisfied if $B^*B=1$.  Given the above properties of the $A$, $B$ and $C$ matrices, we find the simple result
\begin{equation}
    B^*B=\epsilon.
\end{equation}
Therefore, Majorana (or pseudo-Majorana) spinors can only exist if $\epsilon=1$.  For $\epsilon=-1$, the best one can do is to define symplectic-Majorana spinors.  The possible cases are shown in Table~\ref{tbl:maj}.

%%%%%%%%%%%
\begin{table}[t]
\centering
\begin{tabular}{rrcl}
$\epsilon$&$\eta$&$s-t\mod8$&\\
\hline
1&1&0, 1, 2&Majorana\\
1&$-1$&0, 6, 7&pseudo-Majorana\\
$-1$&1&4, 5, 6&symplectic-Majorana\\
$-1$&$-1$&2, 3, 4&pseudo-symplectic-Majorana
\end{tabular}
\caption{Possible sign choices for $\epsilon$ and $\eta$ and the corresponding spinor reality conditions in spacetimes with $t$ time and $s$ space dimensions.  Note the Bott periodicity of real Clifford algebras.}
\label{tbl:maj}
\end{table}
%%%%%%%%%%%%

For a Lorentzian signature with $t=1$, the charge conjugation matrix satisfies
\begin{equation}
    \Gamma_\mu^T=-\eta C\Gamma_\mu C^{-1},\qquad C^T=-\epsilon\eta C\qquad(\mbox{Lorentzian}).
\end{equation}
This corresponds to $t_0=\eta$ and $t_1=-\epsilon$ in the convention of \cite{Freedman:2012zz}.  For Euclidean spinors with $t=0$, we have instead
\begin{equation}
        \Gamma_\mu^T=\eta C\Gamma_\mu C^{-1},\qquad C^T=\epsilon C\qquad(\mbox{Euclidean}).
\end{equation}
With this in mind, we now give explicit realizations of the Dirac matrices for the cases of interest.

%%%%%%%%%%%%%
\subsection{Decomposition of 10 dimensional Majorana-Weyl Spinors for D3 giants}\label{app: Spinors 10 D}
We follow the following conventions and notations of decomposing the $SO(1,9)$ Dirac matrices to $SO(1,3) \times SO(6)$. For the $3+1$ dimensional world-volume, we choose a basis where
\begin{equation}
    \gamma^0 = i\sigma^2 \otimes \mathbf{1}_2, \quad \gamma^i = \sigma^1 \otimes \sigma^i.
\label{eq:Dirac3+1}
\end{equation}
In this basis, following the standard conventions in Lorentzian signature, we define the parity matrix
\begin{equation}
    \gamma^5 = -i\,\gamma^0\gamma^1\gamma^2\gamma^3 = \sigma^3 \otimes \mathbf{1}_2,
\end{equation}
and the Majorana matrix
\begin{equation}
    b_4 =\gamma^5\gamma^2= i\sigma^2\otimes \sigma^2.
\end{equation}
This $b_4$ matrix corresponds to $\epsilon=1$ and $\eta=1$.

For the internal $SO(6)$, we define
\begin{equation}
    \begin{split}
        \tilde\gamma^1 = \sigma^1\otimes\mathbf{1}_2 \otimes\mathbf{1}_2,\nn\\
        \tilde\gamma^2 = \sigma^3\otimes\sigma^1 \otimes\mathbf{1}_2,\nn\\
        \tilde\gamma^3 = \sigma^3\otimes\sigma^3 \otimes\sigma^1, \nn\\
    \end{split}
    \qquad
    \begin{split}
        \tilde\gamma^4 = \sigma^2\otimes\mathbf{1}_2 \otimes\mathbf{1}_2,\nn\\
        \tilde\gamma^5 = \sigma^3\otimes\sigma^2 \otimes\mathbf{1}_2,\nn\\
        \tilde\gamma^6 = \sigma^3\otimes\sigma^3 \otimes\sigma^2.
    \end{split}
\end{equation}
Again following the standard conventions in Euclidean signature, the parity and Majorana matrix are defined respectively as,
\begin{align}
    \tilde\gamma^7 &= -i \tilde\gamma^1 \cdots \tilde\gamma^6 = \sigma^3 \otimes \sigma^3 \otimes \sigma^3\\
    \tilde b_6 &= \tilde\gamma^4 \tilde\gamma^5 \tilde\gamma^6 = i\sigma^2\otimes\sigma^1\otimes\sigma^2.
\end{align}
Here, we have $\epsilon=1$ and $\eta=-1$.  It will be useful to define the two-component spinors
\begin{equation}
    \eta_+ =
    \begin{pmatrix}
        1\\
        0
    \end{pmatrix} ,\qquad
    \eta_- =
    \begin{pmatrix}
        1\\
        0
    \end{pmatrix},
\end{equation}
so that spinors of the form
\begin{equation}
    \eta_{abc} = \eta_a\otimes\eta_b\otimes\eta_c,
\end{equation}
have definite parity under $\tilde\gamma^7$. In our notation $\eta^{(\pm)}$ have $\pm$ parity under $\tilde\gamma^7$.

Combining the D3-brane worldvolume and transverse directions, the $SO(1,9)$ Dirac matrices take the form
\begin{align}
    \Gamma^\mu &= \gamma^\mu \otimes \mathbf{1}_8, \kern2.5em\mu = 0,\cdots,3,\nn\\
    \Gamma^i &= \gamma^5\otimes \tilde\gamma^{i-3}, \qquad i = 4,\cdots,9,\nn\\
    \Gamma^{(11)} &= \Gamma^0 \cdots \Gamma^9 = - \gamma^5 \otimes\tilde\gamma^7,\nn\\
    B_{10} &= \Gamma^2 \Gamma^7 \Gamma^8 \Gamma^9 = - b_4 \otimes \tilde b_6.
\end{align}
In $9+1$ dimensions we have $\epsilon=1$ and $\eta=1$.

%%%%%%%%%%%%%%%%%%%%%%%%%
\subsection{Decomposition of 11 dimensional Majorana Spinors for M5 giants}\label{app: Spinors 11 D M5}

For the decomposition of the $SO(1,10)$ Dirac matrices to $SO(1,5) \times SO(5)$, it is convenient to consider a $1+5+5$ split.  Five-dimensional Euclidean Dirac matrices can be taken to be
\begin{equation}
    \begin{split}
        \tilde\gamma^1 = \sigma^1 \otimes \mathbf1_2,\nn\\
        \tilde\gamma^2 = \sigma^2 \otimes \mathbf1_2,\nn\\
    \end{split} \qquad
    \begin{split}
        \tilde\gamma^3 = \sigma^3 \otimes \sigma^1,\nn\\
        \tilde\gamma^4 = \sigma^3 \otimes \sigma^3,\nn\\
    \end{split}\qquad
    \begin{split}
        \tilde\gamma^5 = \sigma^3 \otimes \sigma^2.\nn\\ 
        \hbox{}
    \end{split}
\end{equation}
Note that $\tilde\gamma^1\cdots\tilde\gamma^5=\mathbf1_2\otimes\mathbf1_2$.  Furthermore, the Majorana matrix is $\tilde b_5=\tilde\gamma^2\tilde\gamma^5=i\sigma^1\otimes\sigma^2$, from which we can read off $\epsilon=-1$ and $\eta=1$.  With this in mind, for the $M_5$ brane world-volume, we choose
\begin{equation}
    \gamma^0=i\sigma^2\otimes\mathbf1_4,\qquad\gamma^i=\sigma^1\otimes\tilde\gamma^i.
\label{eq:Dirac5+1}
\end{equation}
In this case, $\gamma^7=\gamma^0\cdots\gamma^5=\sigma^3\otimes\mathbf1_4$, and the $SO(1,5)$ Majorana matrix is $b_6=\mathbf1_2\otimes\tilde b_5$.  This corresponds to $\epsilon=-1$ and $\eta=1$.

In this decomposition, the $SO(1,10)$ Dirac matrices take the form
\begin{align}
    \Gamma^\mu &= \gamma^\mu \otimes \mathbf{1}_4, \kern2.5em\mu = 0,\cdots,5,\nn\\
    \Gamma^i &= \gamma^7\otimes \tilde\gamma^{i-5}, \qquad i = 6,\cdots,10,\nn\\
    B_{11} &= b_6 \otimes \tilde b_5.
\end{align}
In $10+1$ dimensions we have $\epsilon=1$ and $\eta=1$.  However, we see that the $SO(1,5)$ and $SO(5)$ spinors cannot be taken to be independently Majorana.

\subsection{Decomposition of 11 dimensional Majorana Spinors for M2 giants}\label{app: Spinors 11 D M2}
We follow the following conventions and notations of decomposing the $SO(1,10)$ $\Gamma$ matrices to $SO(1,2) \times SO(8)$. For the world-volume, we choose a basis were,
\begin{equation}
    \gamma^0 = i\sigma^2,\quad \gamma^1 = \sigma^1,\quad \gamma^2 = \sigma^3.
\end{equation}
In this basis the Majorana matrix is simply $b_3 = \mathbf{1}_2$, with corresponding $\epsilon=1$ and $\eta=1$. 

For $SO(8)$ we choose,
\begin{equation}
    \begin{split}
        \tilde\gamma^1 = \sigma^1\otimes\mathbf{1}_2 \otimes\mathbf{1}_2\otimes\mathbf{1}_2,\nn\\
        \tilde\gamma^2 = \sigma^3\otimes\sigma^1 \otimes\mathbf{1}_2\otimes\mathbf{1}_2,\nn\\
        \tilde\gamma^3 = \sigma^3\otimes\sigma^3 \otimes\sigma^1 \otimes\mathbf{1}_2,\nn\\
        \tilde\gamma^4 = \sigma^3\otimes\sigma^3\otimes\sigma^3 \otimes\sigma^1,
    \end{split}
    \qquad
    \begin{split}
        \tilde\gamma^5 = \sigma^2\otimes\mathbf{1}_2 \otimes\mathbf{1}_2\otimes\mathbf{1}_2,\nn\\
        \tilde\gamma^6 = \sigma^3\otimes\sigma^2 \otimes\mathbf{1}_2\otimes\mathbf{1}_2,\nn\\
        \tilde\gamma^7 = \sigma^3\otimes\sigma^3 \otimes\sigma^2\otimes\mathbf{1}_2,\nn\\
        \tilde\gamma^8 = \sigma^3\otimes\sigma^3\otimes\sigma^3 \otimes\sigma^2
.    \end{split}
\end{equation}
In these conventions,
\begin{equation}
    \tilde\gamma^9 = -\sigma^3\otimes \sigma^3\otimes\sigma^3\otimes\sigma^3,
\end{equation}
and
\begin{equation}
    \tilde b_8 = \sigma^1\otimes\sigma^2\otimes\sigma^1\otimes\sigma^2,
\end{equation}
so that $\epsilon=1$ and $\eta=1$.  Similar to the case of $SO(6)$ spinors, one can define $\eta_{abcd}$ as spinors with definite parity under $\tilde\gamma^9$.

In these conventions, the $SO(10,1)$ spinors take the form,
\begin{align}
    \Gamma^\mu &= \gamma^\mu \otimes \tilde\gamma^9, \kern2.6em  \mu = 0,1,2,\nn\\
    \Gamma^i &= \mathbf{1}_2 \otimes \tilde\gamma^{i-2},\qquad i = 3,\ldots,10,\nn\\
    B_{11} &= b_3 \otimes \tilde b_8.
\end{align}
In $10+1$ dimensions, we have $\epsilon=1$ and $\eta=1$.

\section{\texorpdfstring{$U(N)$ and $SU(N)$}{U(N) and SU(N)} characters}
\label{app:characters}

Here we review some basic features of irreducible representations of $SU(N)$ and their corresponding characters.  The irreducible representations of $U(N)$ can be labeled by Young tableau with no more than $N$ rows.  Such diagrams correspond to Dynkin labels $(a_1,a_2,\ldots,a_N)$ where $a_i$ gives the number of columns of height $i$.  For the $SU(N)$ case, columns of height $N$, corresponding to antisymmetrization on $N$ indices can be removed, so irreducible representations of $SU(N)$ are labeled by Young tableau with no more that $N-1$ rows or by the Dynkin labels $(a_1,a_2,\ldots,a_{N-1})$.  The fundamental representation, $\mathbf N$, of $SU(N)$ corresponds to $(100\ldots0)$, while the anti-fundamental representation, $\bar{\mathbf N}$, corresponds to $(0\ldots001)$.

Using the Weyl character formula, $U(N)$ characters can be written as a ratio of determinants
\begin{equation}
    \chi^{U(N)}_{[\vec h]}=\left|\begin{matrix}x_1^{h_1+N-1}&x_1^{h_2+N-2}&\cdots&x_1^{h_N}\\x_2^{h_1+N-1}&x_2^{h_2+N-2}&\cdots&x_2^{h_N}\\\vdots&\vdots&\ddots&\vdots\\x_N^{h_1+N-1}&x_N^{h_2+N-2}&\cdots&x_N^{h_N}\end{matrix}\right|\times\left|\begin{matrix}x_1^{N-1}&x_1^{N-2}&\cdots&1\\x_2^{N-1}&x_2^{N-2}&\cdots&1\\\vdots&\vdots&\ddots&\vdots\\x_N^{N-1}&x_N^{N-2}&\cdots&1\end{matrix}\right|^{-1}.
\end{equation}
Here $[\vec h]=[h_1,h_2,\ldots,h_n]$ where $h_i$ is the width of the $i$-th row of the corresponding Young diagram.  These are related to Dynkin labels by
\begin{equation}
    a_1=h_1-h_2,\quad a_2=h_2-h_3,\quad\ldots,\quad a_{N-1}=h_{N-1}-h_N,\qquad a_N=h_N.
\end{equation}
The related $SU(N)$ characters are obtained by taking $x_1x_2\ldots x_N=1$ and by dropping $a_N$ from the Dynkin labels.

As an example, the $U(2)$ character is given by
\begin{equation}
    \chi^{U(2)}_{[h_1,h_2]}=\fft{\left|\begin{matrix}x_1^{h_1+1}&x_1^{h_2}\\x_2^{h_1+1}&x_2^{h_2}\end{matrix}\right|}{\left|\begin{matrix}x_1&1\\x_2&1\end{matrix}\right|}=(x_1x_2)^{\fft{h_1+h_2}2}\fft{(x_1/x_2)^{\fft{h_1-h_2+1}2}-(x_1/x_2)^{-\fft{h_1-h_2+1}2}}{(x_1/x_2)^{\fft12}-(x_1/x_2)^{-\fft12}}.
\end{equation}
Rewriting this in terms of Dynkin labels $(a_1,a_2)$ gives
\begin{equation}
    \chi^{U(2)}_{(a_1,a_2)}=(x_1x_2)^{\fft{a_1+2a_2}2}\fft{(x_1/x_2)^{\fft{a_1+1}2}-(x_1/x_2)^{-\fft{a_1+1}2}}{(x_1/x_2)^{\fft12}-(x_1/x_2)^{-\fft12}}.
\end{equation}
Note that $a_1+2a_2$ is the total number of boxes in the Young diagram.  Taking $x_2=1/x_1$ gives the corresponding $SU(2)$ character
\begin{align}
    \chi^{SU(2)}_{(a)}=\fft{x_1^{a+1}-x_1^{-(a+1)}}{x_1-x_1^{-1}}&=x_1^{a}+x_1^{a-2}+\cdots+x_1^{2-a}+x_1^{-a}\nn\\
    &=x_1^{a}+x_1^{a-1}x_2+x_1^{a-2}x_2^2+\cdots+x_2^{a}.
\end{align}

The expansion of the determinants for the $SU(4)$ characters that show up in the M5 giant graviton fluctuations gives rise to a rather lengthy expression.  However, we note that the sum over scalar harmonics on $S^5$ that arises in (\ref{eq:su4fscalar}) can be evaluated to give
\begin{equation}
    \sum_{\ell\ge0}z^{\ell/2}\chi^{SU(4)}_{(0,\ell,0)}=\fft{1-z}{\prod_{1\le i<j\le4}(1-\sqrt{z}\,x_ix_j)}.
\end{equation}
Similarly, we find
\begin{equation}
    \sum_{\ell\ge1}z^{\ell/2}\chi^{SU(4)}_{(0,\ell-1,2)}=\fft{z^{1/2}\chi^{SU(4)}_{(0,0,2)}-  z\chi^{SU(4)}_{(1,0,1)}+z^{3/2}\chi^{SU(4)}_{(0,1,0)}-z^2}{\prod_{1\le i<j\le4}(1-\sqrt{z}\,x_ix_j)}.
\end{equation}
Decomposing the $SU(4)$ characters into $SU(3)$ characters with the mapping
\begin{equation}
    x_1=\left(\fft{\sqrt{z}}\lambda\right)^{-3/2},\qquad x_i=\left(\fft{\sqrt{z}}\lambda\right)^{1/2}\left(\fft{y_i}\lambda\right),\qquad i=2,3,4,
\end{equation}
with $\lambda=(y_2y_3y_4)^{1/3}$ then gives the tensor contribution, (\ref{eq:IM5tensor}).  Finally, the fermion harmonics give rise to the sum
\begin{equation}
    \sum_{\ell\ge0}z^{\ell/2}\chi^{SU(4)}_{(0,\ell,1)}=\fft{\chi^{SU(4)}_{(0,0,1)}-\sqrt{z}\chi^{SU(4)}_{(1,0,0)}}{\prod_{1\le i<j\le4}(1-\sqrt{z}\,x_ix_j)},
\end{equation}
which can similarly be decomposed to yield (\ref{eq:IM5fermion}).

%%%%%%%%%%%%%%%%%%%%%%%%%%%%%%%%%%%%%%%%%%%

\section{Fluctuations from non-abelian DBI}\label{app:nA DBI}
The scalar fluctuations of a stack of D3 branes are described by the non-abelian generalization of DBI in \cite{Myers:2003bw}. In a background with metric $g_{\mu\nu}$, constant dilaton, and no Kalb-Rammond field, the the action for the matrix-valued scalars $\Phi^i$ consists of two terms,
\begin{align}\label{eq:nA DBI}
    S=-T \int& d^4 \sigma \operatorname{STr}\left( \sqrt{\operatorname{det}\left(Q^i{ }_j\right)}\sqrt{-\operatorname{det}\left(P\left[g_{\mu \nu}+g_{\mu i}\left(Q^{-1}-\delta\right)^{i j} g_{j \nu}\right]\right)}\right)\nonumber\\
    &+T\int \operatorname{STr}\left(P\left[e^{i \lambda \mathrm{i}_{\Phi} \mathrm{i}_{\Phi}}\left(\sum C^{(n)}\right)\right]\right),
\end{align}
where $Q^i{ }_j \equiv \delta^i{ }_j+i \lambda\left[\Phi^i, \Phi^k\right] g_{k j}$ and
\begin{equation}\label{eq:iiC}
    \mathrm{i}_{\Phi} \mathrm{i}_{\Phi} C^{(n)}=\frac{1}{2(n-2)!}\left[\Phi^i, \Phi^j\right] C_{j i \mu_3 \cdots \mu_n}^{(n)} d x^{\mu_3} \cdots d x^{\mu_n}.
\end{equation}
Greek indices denote all coordinates, and Latin indices are split with $i,j,k\dots$ perpendicular to the brane and $a,b,c\dots$ along the worldvolume. The determinants are taken over spacetime indices and the symmetrized trace over gauge indices. Expanding the first determinant term in \eqref{eq:nA DBI}  in $\Phi$ gives
\begin{equation}
\sqrt{\operatorname{det}\left(Q^i{ }_j\right)}=1-\fft{\lambda^2}{4}[\Phi^i,\Phi^j][\Phi_i,\Phi_j]+\mathcal{O}(\lambda^4).
\end{equation}
Quadratic terms disappear because of the antisymmetry of the commutator. In the other determinant we have a pullback, which for a general tensor $E_{\mu\nu}$ is given by
\begin{equation}
    P[E]_{a b}=E_{a b}+\lambda E_{a i} D_b \Phi^i+\lambda E_{i b} D_a \Phi^i+\lambda^2 E_{i j} D_a \Phi^i D_b \Phi^j+\mathcal{O}(\lambda^3).
\end{equation}
For a metric without cross terms between worldvolume and transverse direction (i.e. $g_{ai}=0$) the pullback simplifies leading to an absence of commutator terms quadratic and quartic in $\Phi$:
\begin{align}
    P\left[g_{\mu \nu}+g_{\mu i}\left(Q^{-1}-\delta\right)^{i j} g_{j \nu}\right]_{ab}&=P\left[g_{\mu\nu}+g_{\mu i}\left(-i\lambda[\Phi^i,\Phi^j]+\lambda^2[\Phi^i,\Phi^k][\Phi_k,\Phi^j]\right)g_{j\nu}+\cdots\right]_{ab}\nonumber\\
    &=g_{ab}+\lambda^2 g_{i j} D_a \Phi^i D_b \Phi^j+\mathcal{O}(\lambda^4)
\end{align}
For the Wess-Zumino term, the scalar commutator in \eqref{eq:iiC} will never appear. This is because only $C^{(4)}$ is nonzero, and $[\Phi^i,\Phi^j]$ cannot be contracted without reducing the degree of the form. Therefore
\begin{equation}
    S_{W Z}=T\int \operatorname{STr}\left(P\left[C^{(4)}\right]\right).
\end{equation}
Putting all the terms together gives
\begin{align}\label{eq:expanded_BI}
    S=-T\int d^{4}& \sigma \operatorname{STr}\left[ \sqrt{-\operatorname{det}\left(g_{ab}\right)}\left(1+\fft{\lambda^2}{2}g^{ab}g_{ij}D_a\Phi^i D_b\Phi^j-\fft{\lambda^2}{4}[\Phi^i,\Phi^j][\Phi_i,\Phi_j]+\mathcal{O}(\lambda^4)\right)\right]\nn\\
    &+T\int \operatorname{STr}\left(P\left[C^{(4)}\right]\right).
\end{align}
To continue with the fluctuation analysis of this action, one would perform a non-Abelian Taylor expansion of the metric. However, all non-abelian effects (beyond a simple promotion of fields to matrices) only appear past quadratic order.

%%%%%%%%%%%%%%%%%%%%
\bibliographystyle{JHEP}
\bibliography{refs}

\end{document}